\documentclass{iopart}

\usepackage{iopams}
\usepackage{graphicx}
\usepackage{color}
\usepackage{comment}
\usepackage{subfigure}
\usepackage{inputenc}

\newcommand{\beq}{\begin{equation}}
\newcommand{\eeq}{\end{equation}}
\newcommand{\bdm}{\begin{displaymath}}
\newcommand{\edm}{\end{displaymath}}

\begin{document}

\title[Seismic array measurements at Virgo's WEB]{Seismic array measurements at Virgo's West End Building for the configuration of a Newtonian-noise cancellation system}

\author{Maria C. Tringali, Tomasz Bulik}
\vskip 1mm
\address{Astronomical Observatory Warsaw University, 00-478 Warsaw, Poland}
\author{Jan Harms}
\address{Gran Sasso Science Institute (GSSI), I-67100 L'Aquila, Italy}
\address{INFN, Laboratori Nazionali del Gran Sasso, I-67100 Assergi, Italy}
\author{Irene Fiori}
\vskip 1mm
\address{European Gravitational Observatory (EGO), I-56021 Cascina, Pisa, Italy}
\author{Federico Paoletti}
\address{INFN, Sezione di Pisa, I-56127 Pisa, Italy}
\author{Neha Singh, Bartosz Idzkowski, Adam Kutynia, Krzysztof Nikliborc, Maciej Suchi\'{n}ski}
\vskip 1mm
\address{Astronomical Observatory Warsaw University, 00-478 Warsaw, Poland}
\author{Alessandro Bertolini, Soumen Koley}
\vskip 1mm
\address{Nikhef, Science Park, 1098 XG Amsterdam, Netherlands}

\begin{abstract}

Terrestrial gravity fluctuations produce so-called Newtonian noise (NN) which is expected to limit the low frequency sensitivity of existing gravitational--waves (GW) detectors LIGO and Virgo, when they will reach their full potential, and of next--generation detectors like the Einstein Telescope.
In this paper, we present a detailed characterization of the seismic field at Virgo's West End Building as part of the development of a Newtonian noise cancellation system. The cancellation system will use optimally filtered data from a seismometer array to produce an estimate of the Newtonian--noise generated by the seismic field, and to subtract this estimate from the gravitational--wave channel of the detector. By using an array of 38~seismic sensors, we show that, despite the influence of the complexity of Virgo's infrastructure on the correlation across the array, Wiener filtering can still be very efficient in reconstructing the seismic field around the test--mass location. Taking into account the division of the building's foundations into separate concrete slabs, and the different properties of the seismic field across them, we conclude that the arrays to be used for the Newtonian--noise cancellation at Virgo will require a relatively large number of seismometers per test mass, i.e. significantly more than~10. Moreover, observed variations of the absolute noise residuals over time, related to the daily evolution of anthropogenic noise, suggest that the Wiener filter will need to be updated regularly, probably more often than every hour, to achieve stationarity of the background level after subtraction.

\end{abstract}


\section{Introduction}

Terrestrial gravity fluctuations, known as Newtonian noise (NN), were first predicted to be a limiting noise source for ground-based gravitational-wave (GW) detectors by Weiss \cite{Wei1972}, and first analyzed in greater detail by Saulson \cite{Saulson}. The sources of NN are mass density fluctuations of both atmospheric and seismic origin.
Atmospheric~NN is produced by pressure fluctuations (infrasound noise), and advection of temperature and humidity fields. A recent dedicated observation campaign performed at the Virgo site \cite{Donatella} highlighted the relevance of mitigating the infrasound noise inside the buildings, for the future upgrades of Advanced Virgo (AdV), and, in the case of next generation underground detectors, inside the cavities housing the test masses.
Seismic NN is generated by compression of the ground medium or by surface and interface displacement.
The main types of seismic waves relevant to NN are compressional waves, shear waves, and Rayleigh waves \cite{Jan2015}. Compressional waves are longitudinal waves producing displacement along the direction of propagation, while the nature of shear waves is transversal since they generate displacement perpendicular to the direction of propagation. Both wave types, compressional and shear, are body waves since they can propagate through media in all directions. Rayleigh waves travel instead along the surface of media producing elliptical motion of the ground particles; their amplitude decreases exponentially as the distance from the surface increases. 
Newtonian noise from seismic fields is expected to become a limiting noise source below 20~Hz for Advanced LIGO \cite{LSC2015} and Advanced Virgo \cite{AcEA2015}, when the detectors will reach their full potential.\\
As reported in \cite{Jan2016,Cel2000,coughlin2018,coughlin2014}, the conventional approach to mitigate this kind of noise is based on Wiener filtering. In general, Wiener filters are used to reduce the variance of data in a target channel by exploiting correlations with auxiliary channels. In the context of NN reduction in GW detectors, the auxiliary channels are environmental seismic sensors while the target channel is the interferometer output.\\ 
In this work, we report the results of a characterization of the seismic field performed inside the West End Building (WEB) of the Virgo interferometer in February 2018 by means of an array of seismic sensors. The collected data have been used to investigate the efficiency of the Wiener filtering method. In particular, we have investigated efficient geometrical configuration of the array, and the stability of the method with time. The paper is organized as follows. In Section~\ref{sec:WFtheory}, we introduce the Wiener filter for single--input single--output and multiple--input single--output noise cancellation. In Section~\ref{sec:WEBsetup}, we give a brief description of the seismometer array setup. Section~\ref{sec:Results} collects the main results of Wiener filtering study. Finally, in Section~\ref{sec:Conclusions}, we report the conclusion of this work.

\section{Wiener filtering theory}
\label{sec:WFtheory}

In this section, we summarize the equations defining a Wiener filter in frequency domain. We consider two implementations: the single-input single-output, and the multi-input single-output filter, which is the relevant one for the purpose of this work. The argumentation of this topic is based on \cite{book1}\cite{book2}.\\
In the following section, we will use the terms target channel and input channels. In the context of Wiener filtering to GW detectors, an array of seismometers on site can be used as input channels and, the target channel is the output data of GW detector. In our work, the target channel is the signal from a single seismometer (see Section \ref{sec:WF}).\\

\subsection{Single-input Single-Output (SISO) filter}
Wiener filters are linear, which means that they are applied as simple convolutions in time domain. 
Let $X(f)$ be the target and $S(f)$ the input signal of the filter from a single auxiliary channel in frequency domain. Since a convolution in time domain is described by a simple multiplication of Fourier amplitudes in frequency domain, the filter output, $\hat{X}(f)$, is given by: 
\begin{equation}
\hat{X}(f)= W(f)S(f)~,
\label{freqWiener}
\end{equation}
where $W(f)$ is filter response. The Wiener filter is the optimal linear filter since it minimizes the  mean--squared error signal $E[|e(f)|^{2}]$ if all data are stationary. The estimation error is given by 
\begin{equation}
    e(f)= X(f) - \hat{X}(f)= X(f) -W(f)S(f) ~ , 
\end{equation}
and the mean--square estimation error at a frequency $f$ can then be written 
\begin{eqnarray}
\mathit{E}\big[|e(f)|^{2}\big] &= \mathit{E}\big\{[X(f) -W(f)S(f)]^{\ast} [X(f) -W(f)S(f)]\big\} 
\label{eq:err_SISO}
\end{eqnarray}
where $\mathit{E}[\cdot]$ is the expectation value and symbol $\ast$ denotes the complex conjugate. The filter producing the least mean-square error is obtained by setting to zero the derivative of this last equation with respect to the filter $W(f)$:
\begin{equation}
  \frac{\partial}{\partial W(f)}\mathit{E}\big[|e(f)|^{2}\big] = 0 
  \qquad \longrightarrow   \qquad   W(f)=C_{SS}(f)^{-1}C_{XS}(f)~,
    \label{eq:WF_siso}
\end{equation}
where $C_{SS}(f)=\mathit{E}[|S(f)|^2]$ is the power spectrum of the input signal $S(f)$ and $C_{XS}(f)=\mathit{E}[X(f)S^{\ast}(f)]$ is the cross-power spectrum between input and target signal $X(f)$. 

\subsection{Multiple-input Single-output (MISO) filter}
In this section, we briefly describe the case of a multiple-input single-output (MISO) filter. Formally, the equations are the same as in the previous section except for the fact that vectors and matrices of cross--power spectra are now introduced. Let's consider a system with $M$ input signals and a single output. Similarly to the SISO filter, the input--output relation becomes
\begin{equation}
    \vec{\hat{X}}(f)= \vec{W}(f)^{T}\cdot\vec{S}(f),
\end{equation}
and the MISO Wiener filter equation is
\begin{equation}
     \vec{W}(f)= {\mathbf C}_{SS}(f)^{-1}\cdot\vec{C}_{XS}(f),
    \label{eq:WF_miso}
\end{equation}

\noindent
where ${\mathbf C}_{SS}(f)= \mathit{E}[\vec{S}(f)\vec{S}^{\dagger}(f)]$ is the $M\times M$ matrix of cross spectra between the $M$ input signals ($\dagger$ denotes the complex-conjugate transposition), and $\vec{C}_{XS}(f)= \mathit{E}[X(f)\vec{S}^{\ast}(f)]$ is an $M$-component vector of cross spectra between the $M$ input signals and the target signal.

\section{West End Building seismometer setup}
\label{sec:WEBsetup}

\begin{figure}[ht!]
    \centering
    {\includegraphics [width=0.45\textwidth]{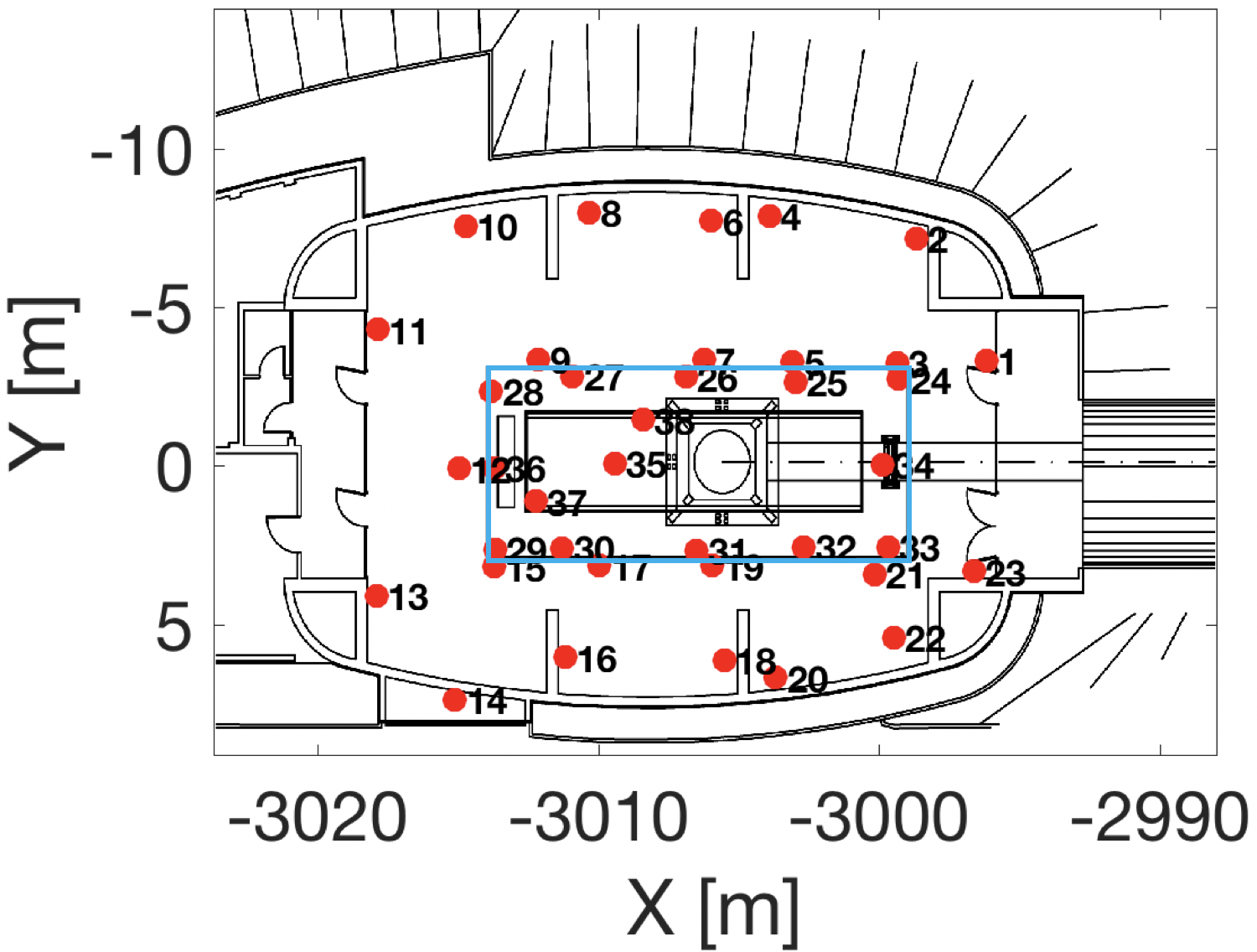}}
    {\includegraphics [width=0.45\textwidth]{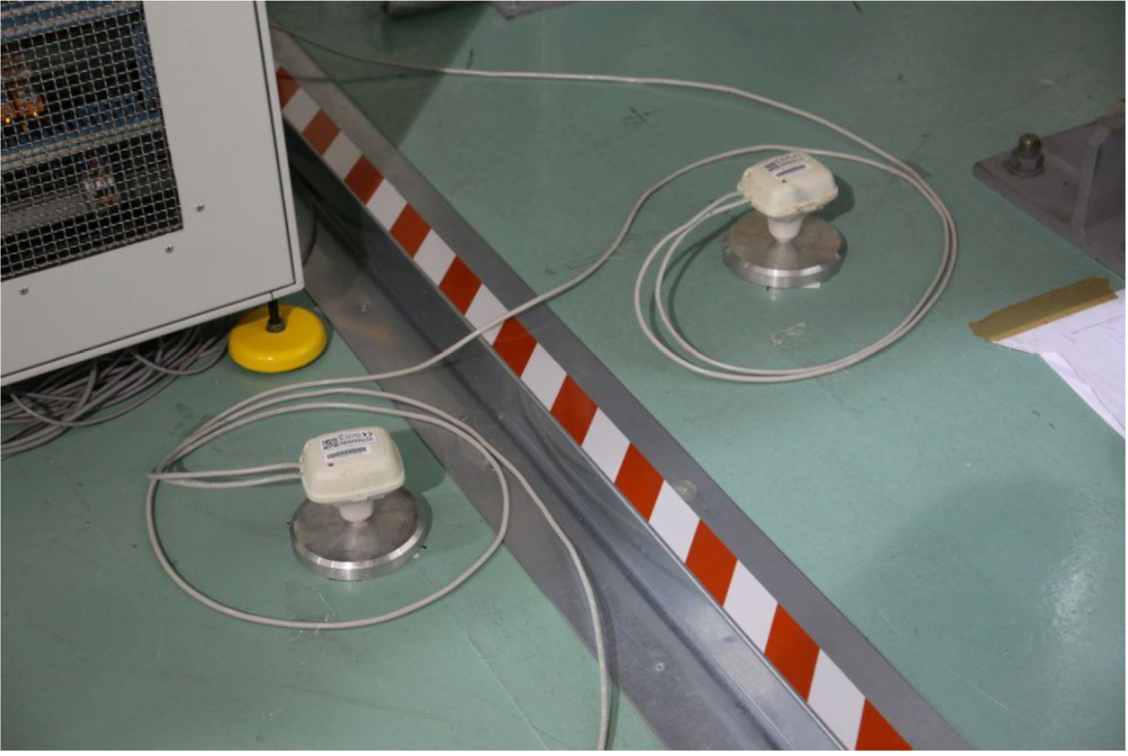}}
    \caption{(Left plot) A map of the seismometer positions at West end building at Virgo site. Light blue rectangular perimeter denotes the extent of the tower platform. (Right plot) Two sensors located near (left) and on tower (right) platform, respectively.}
    \label{fig:WEB_array}
\end{figure}

The Virgo detector is Michelson laser interferometer with 3--km long Fabry--Perot arm cavities that responds to distance changes between suspended test masses caused by passing gravitational waves. Two end buildings, the North and West End Buildings (NEB, WEB), identical in construction and layout, host the two end test masses of the interferometer, while the central building (CEB) hosts the two input test masses.
The foundations of the WEB (and NEB) consists of two concrete slabs, the building platform and the tower platform (see left plot of Figure \ref{fig:WEB_array}). The tower platform carries the vacuum chamber hosting the test--mass suspension and vibration isolation system, the so--called superattenuator \cite{AcEA2015}. Since the Virgo detector area consists of soft soil, the slabs need to be supported by concrete poles several tens of meters long to connect the construction to a deeper, stiffer layer of the ground. The main reason why the foundations are separated into two slabs is cost reduction, since the requirement on the long term subsidence of the tower platform is much stricter than for the rest of the building. Accordingly, the outer slab is supported by 30~m long poles, while the tower floor, together with its basement, is supported by a set of longer poles reaching a more stable gravel layer, which, at the WEB location, is at 52\,m depth. To date, a step of almost 9\,cm has formed between building and tower platforms, as shown in the right plot of Figure \ref{fig:WEB_array}.
Since building and tower platforms are disconnected by a gap about 1\,cm wide and 3.50\,m deep, the construction also provides suppression of seismic noise on the tower platform above 15\,Hz as it will be shown later. 

The tower platform or inner slab constructed in the central part of the building floor is 6\,m wide and 15\,m long. It supports the vacuum tank (the so-called {\it tower}) and the clean room (20\,m$^{2}$) used for payload installation and located at the basement level. The middle part of the tower floor on ground level is covered by a thin metal sheet that can be removed to allow the payload insertion. The building platform or outer slab supports the building's structure i.e. the walls and the roof. The height of the building is 17\,m and the covered area is 17\,m wide and 25\,m long in the arm direction. The back side of each building (towards negative $X$ values in Figure \ref{fig:WEB_array}) hosts a technical room with various electro--mechanical infrastructure devices like the 15\,kV power supply, transformers, uninterrupted power supplies, a diesel generator and chilled/hot water generators for air conditioning \cite{Virgo}. The technical room contains some of the major sources of seismic disturbances in the NN band at Virgo.\\

On January 18 and 19, 2018, an array of 38 seismometers was deployed inside the WEB \cite{Paoli}. Figure \ref{fig:WEB_array} shows the positions of indoor sensors: 14 are located near the walls of the WE tower, 9 near the tower platform, 13 on the tower platform and 2 in the basement of the tower platform. The sensors, manufactured by InnoSeis \cite{arrayNikhef}, are based on 5\,Hz geophones and monitor the vertical ground velocity. The sensor package also contains a pre--amplifier and an analog--to--digital converter to avoid issues with excess electromagnetic (EM) noise coupling when transmitting analog signals through several meter long cables in an EM noisy environment.

The sensors were placed on the floor fixing their heavy mount plate with double--sided adhesive tape for a good connection to ground (see the right plot of Figure \ref{fig:WEB_array}). The data acquisition covers 13 days, from January 25 to February 6, 2018 \cite{data_indoor}. A central data--acquisition unit used for the readout of the entire array also had the purpose of synchronizing and powering the sensors. The sampling frequency of the sensors was set to 500\,Hz.

\begin{figure}[ht!]
    \centering
    {\includegraphics [width=0.49\textwidth]{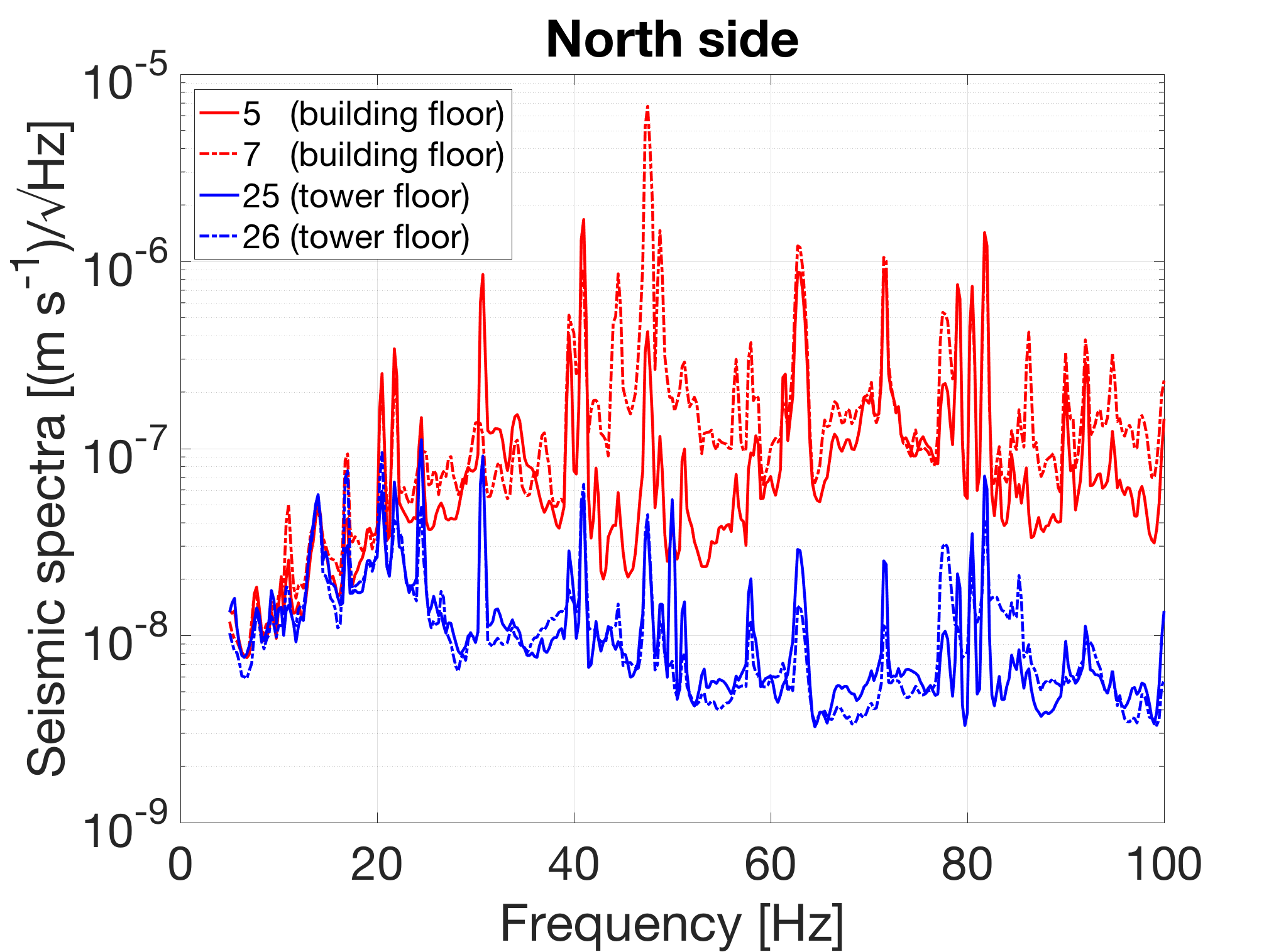}}
    {\includegraphics [width=0.49\textwidth]{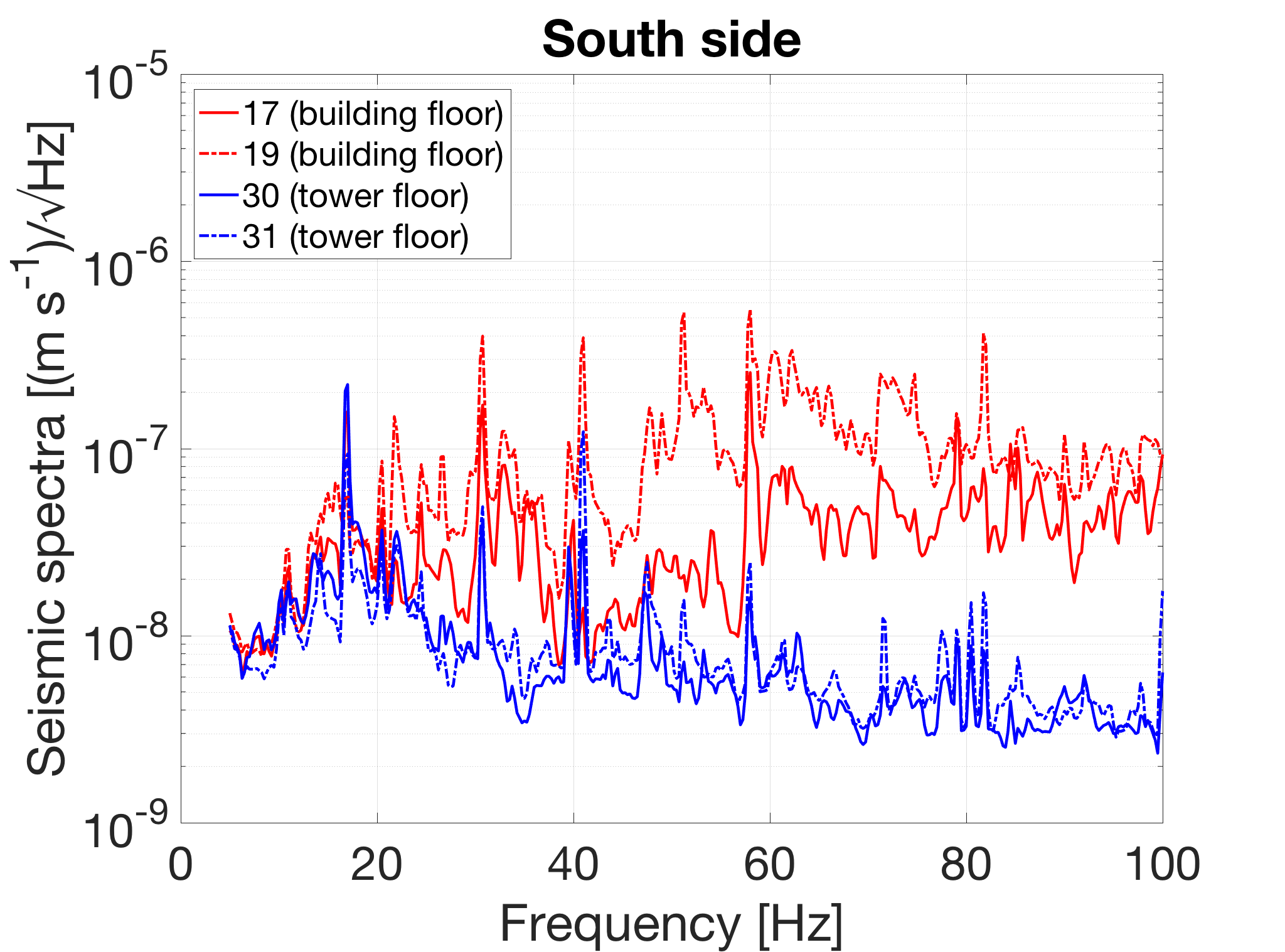}}
   \caption{Seismic spectra of two couples of seismometers positioned at North side and other two at South side of West end building, respectively. Sensors 5, 7, 17, 19 (solid and dashed red curves) lies on building platform while sensors 25, 26 30, 31 (solid and dashed blue curves) on tower platform.}
    \label{fig:WEBsides}
\end{figure}

In Figure \ref{fig:WEBsides}, we report the amplitude spectral density (ASD) of the seismic field measured by four seismometers positioned on the North and South sides of the building. The seismometers are compared in pairs: one sensor is located on the building platform and the other one on the tower platform with a distance of less than 0.5\,m between them. For frequency values above $\sim$ 15\,Hz, a significant difference in ASDs is evident when comparing tower platform (blue curves) and building platform (red curves).

\begin{figure}[ht!]
    \centering
    {\includegraphics [width=0.7\textwidth]{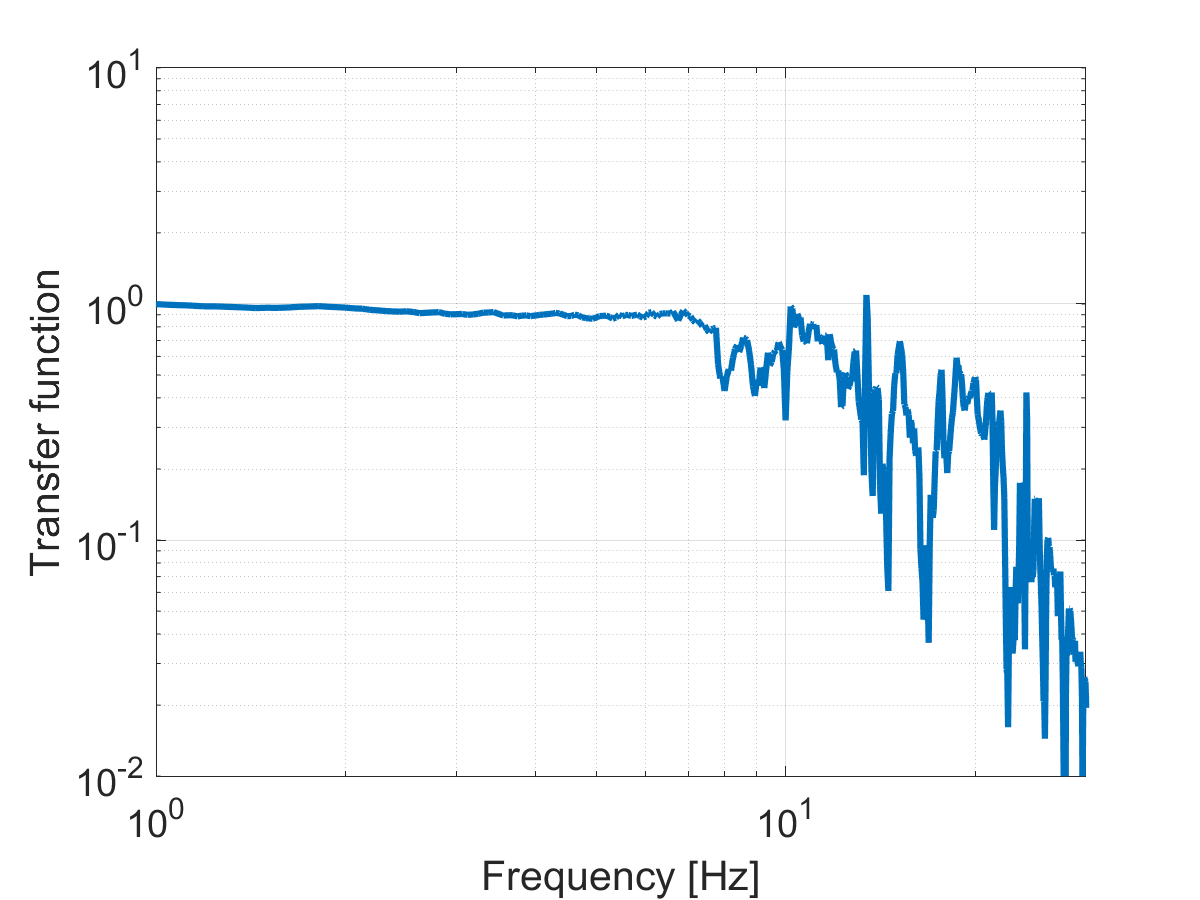}}
   \caption{Absolute value of transfer function from seismometer on building floor to seismometer on tower platform. The distance between the two seismometers is less than 0.5\,m.}
    \label{fig:gaptransfer}
\end{figure}

The explanation for this difference in ASDs is that dominant seismic sources are located outside the tower platform, and that the wavelength of the seismic waves they produce is sufficiently short above 15\,Hz that they get reflected from the gap between the platforms. In other words, the gap does not reach deep enough into the soil to efficiently reflect seismic waves with frequency below 15\,Hz. This effect can be characterized in detail by calculating the transfer function between such a pair of seismometers, as shown in Figure \ref{fig:gaptransfer}.

\section{Wiener filtering investigation}
\label{sec:Results}

In this section of the paper, we present the first results of a Wiener--filtering study. We consider, as case studies, the first hour (UTC) of February 4 and 5, 2018, and we focus on the 1 -- 50\,Hz frequency band.

\begin{figure}[t]
\begin{center}
\includegraphics [width=0.49\textwidth]{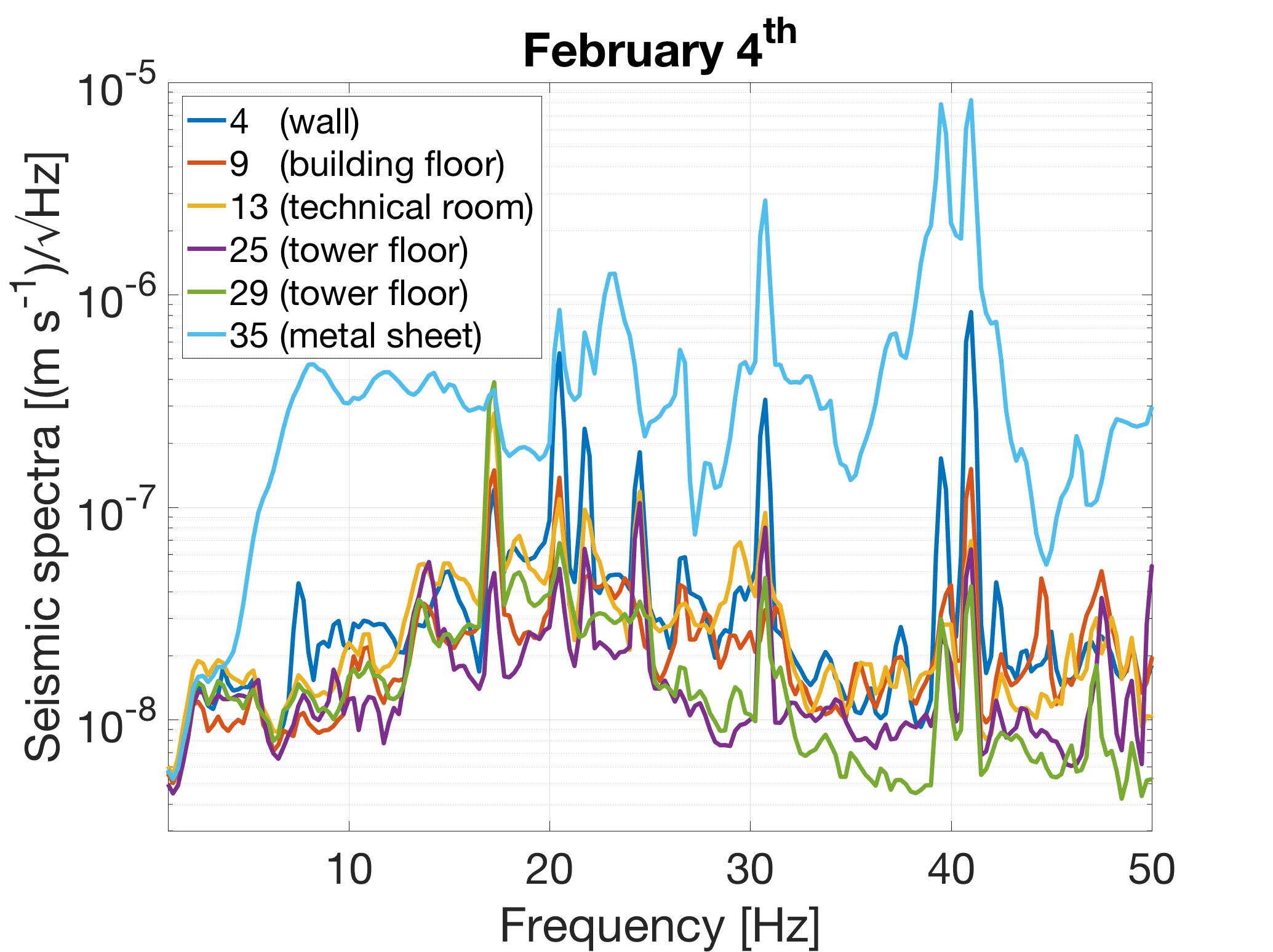}
\includegraphics [width=0.49\textwidth]{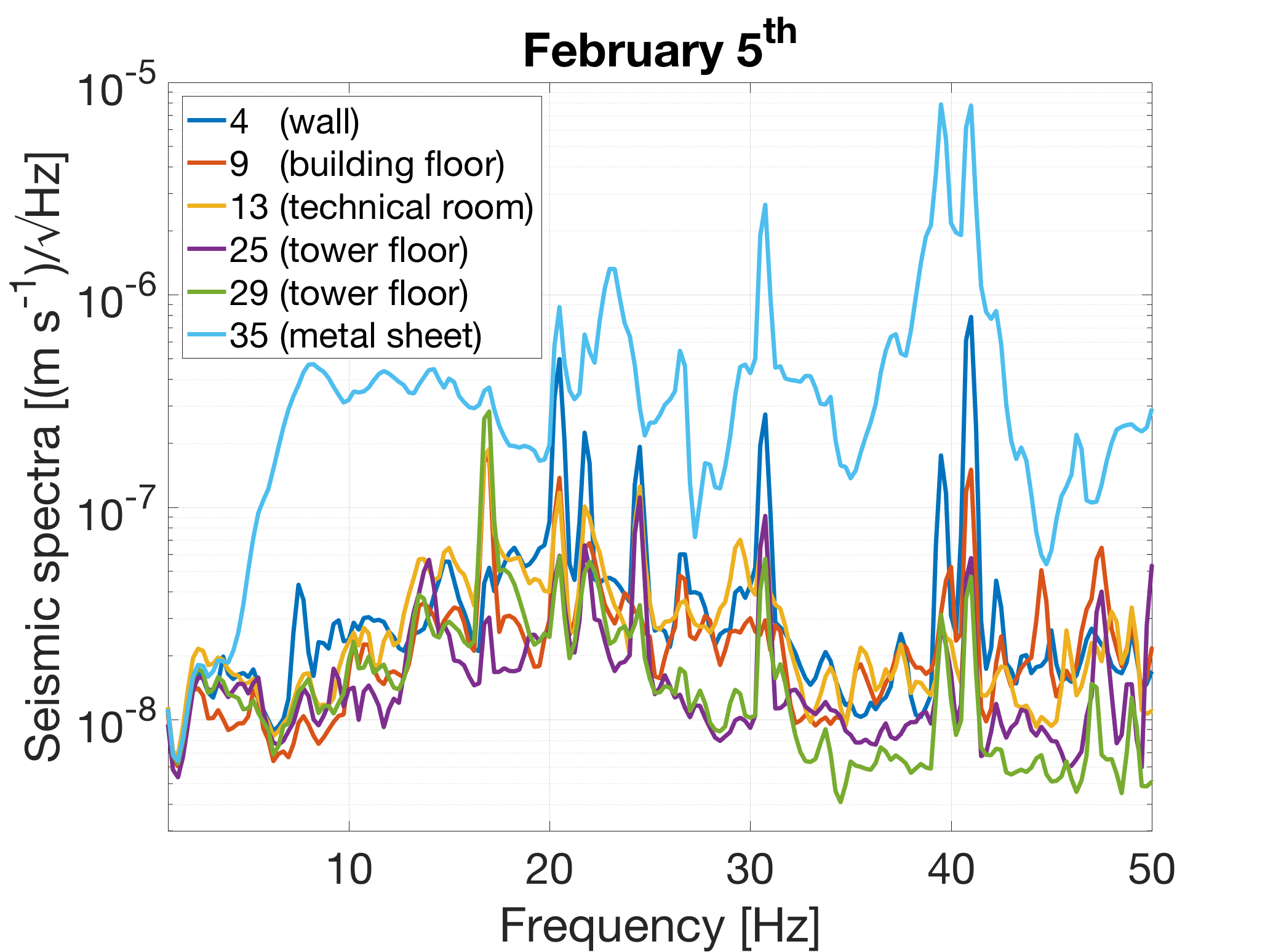}
\caption{Amplitude spectral densities of few seismometers. The spectra are computed on one hour for February 4$^{th}$ (left plot) and 5$^{th}$ (right plot).}
\label{fig:spectra}
\end{center}
\end{figure}

\subsection{Virgo WEB seismometer array}

We first present a characterization of the seismic field in the WEB. 
In the analysis we have down--sampled the data to 250\,Hz, and then divided them into 4\,s (1000 samples) segments. Spectral and cross spectral power densities have been estimated by averaging over the segments by using a Hann window with 50\% overlap. In Figure \ref{fig:spectra}, we report seismic spectra of a few selected seismometers located at the wall of the building (sensor $\#$4), near the technical room (sensor $\#$13), near the tower platform (sensor $\#$9), on the building floor (sensors $\#$25 and $\#$29), and on the metal sheet (sensor $\#$35). The calculation of amplitude spectral densities covers one hour starting from midnight (local time). The data from seismometer 35 clearly stands out above the level of all other units. This is due to the fact that it was positioned on the metal sheet on the tower platform, which vibrates much more strongly than ground probably due to acoustic perturbation. The seismic spectra lie approximately between $10^{-8}$\,m\,s$^{-1}$Hz$^{-1/2}$ and $10^{-7}$\,m\,s$^{-1}$Hz$^{-1/2}$. There are several lines in the spectra that affect the entire building. In particular one can see vibrations at approximately 17\,Hz, 21\,Hz, 22\,Hz, 24\,Hz, 31\,Hz, 39\,Hz ad 41\,Hz. The lines are persistent, and some were identified with disturbances from vacuum pumps and cooling fans.

\begin{figure}
\begin{center}
\includegraphics [width=0.49\textwidth]{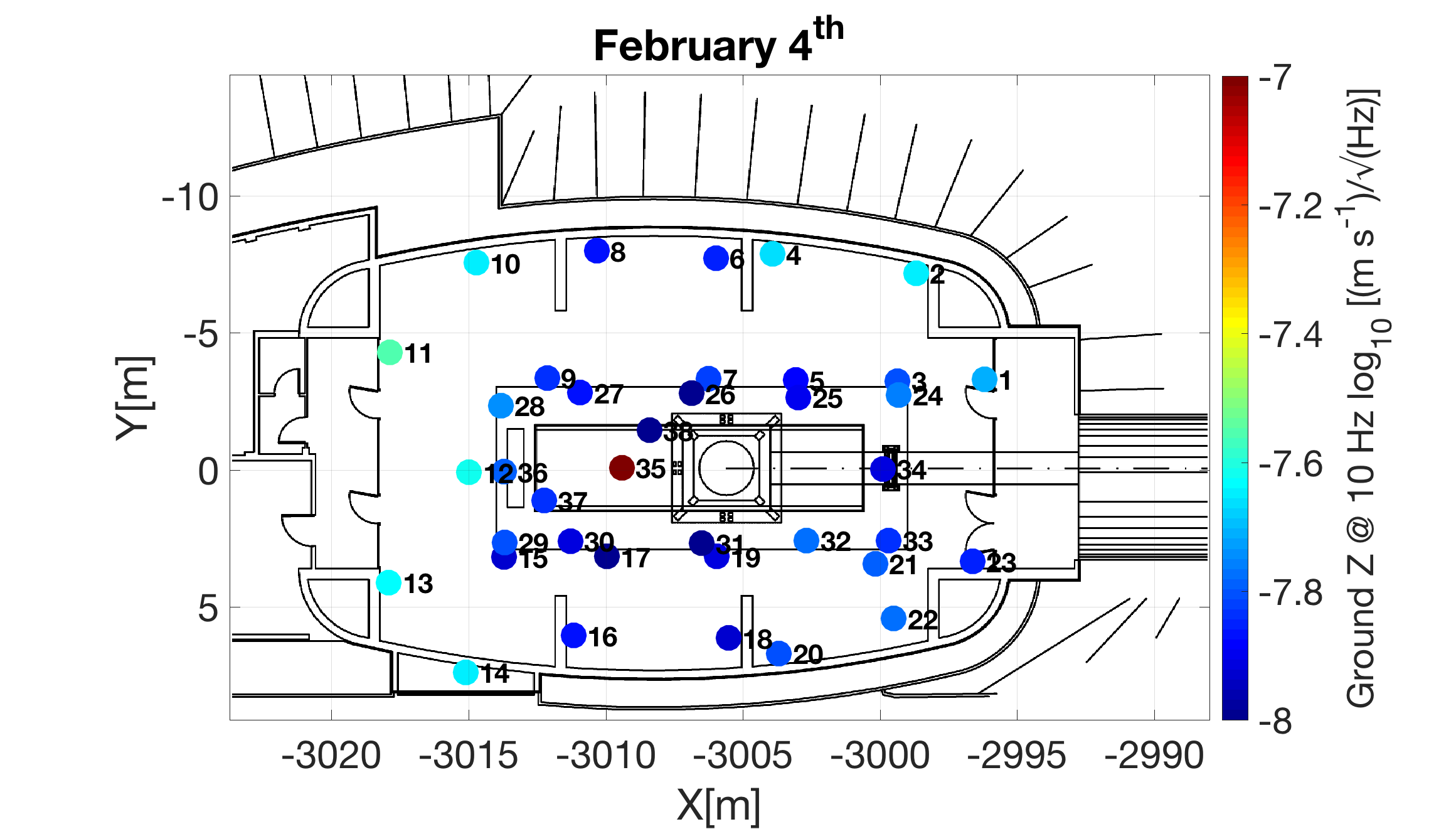}
\includegraphics [width=0.49\textwidth]{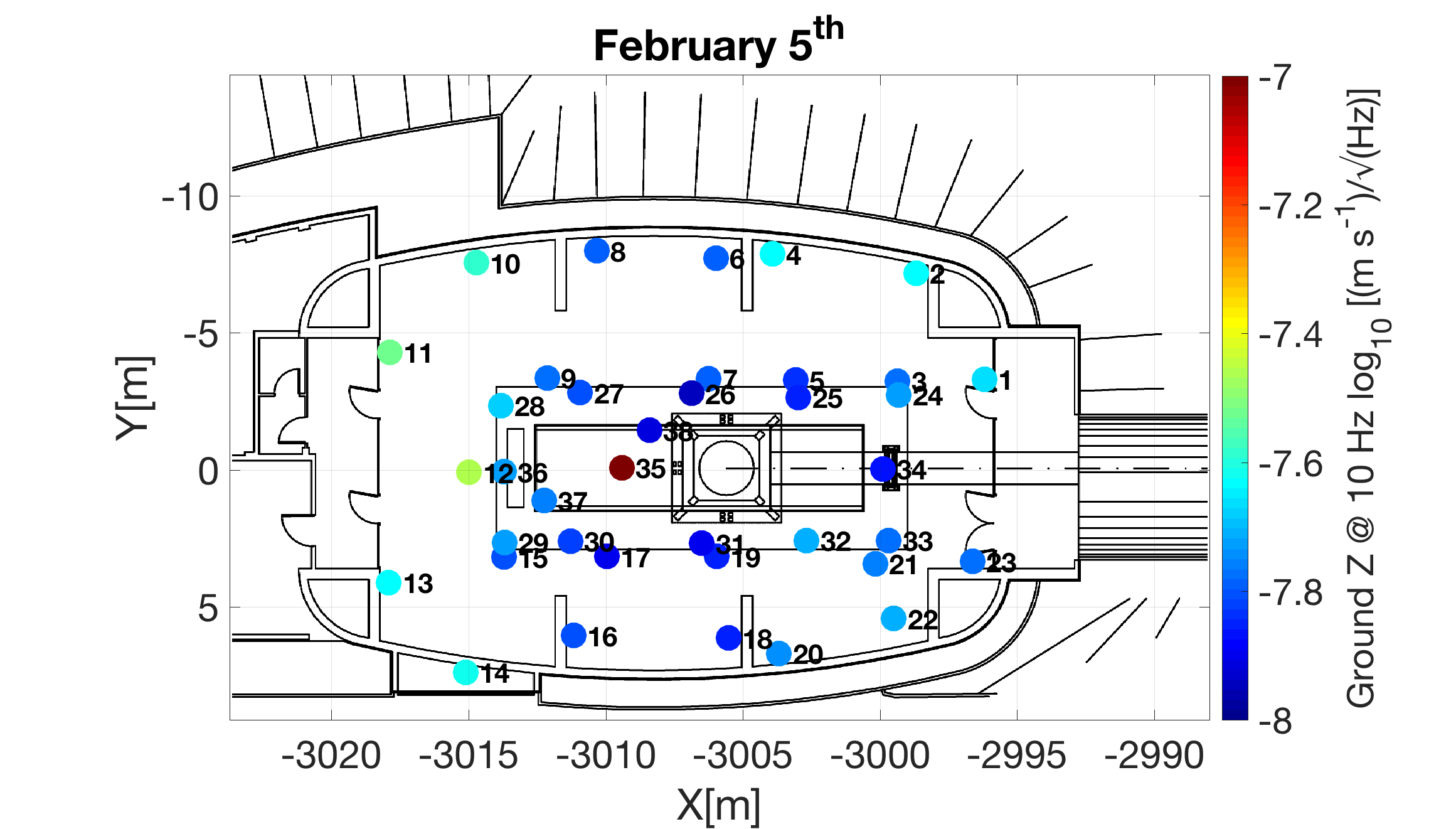}\\

\includegraphics [width=0.49\textwidth]{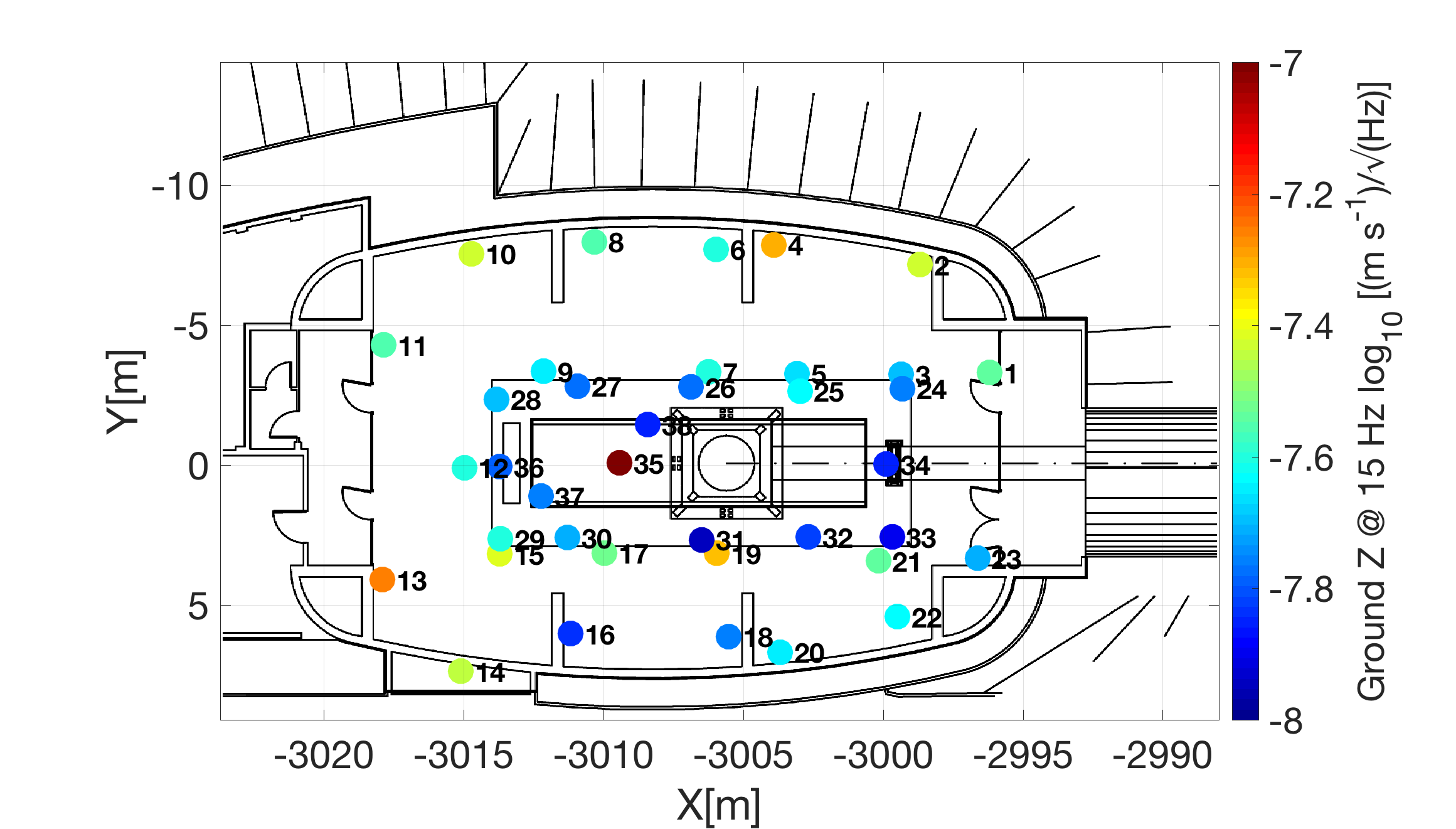}
\includegraphics [width=0.49\textwidth]{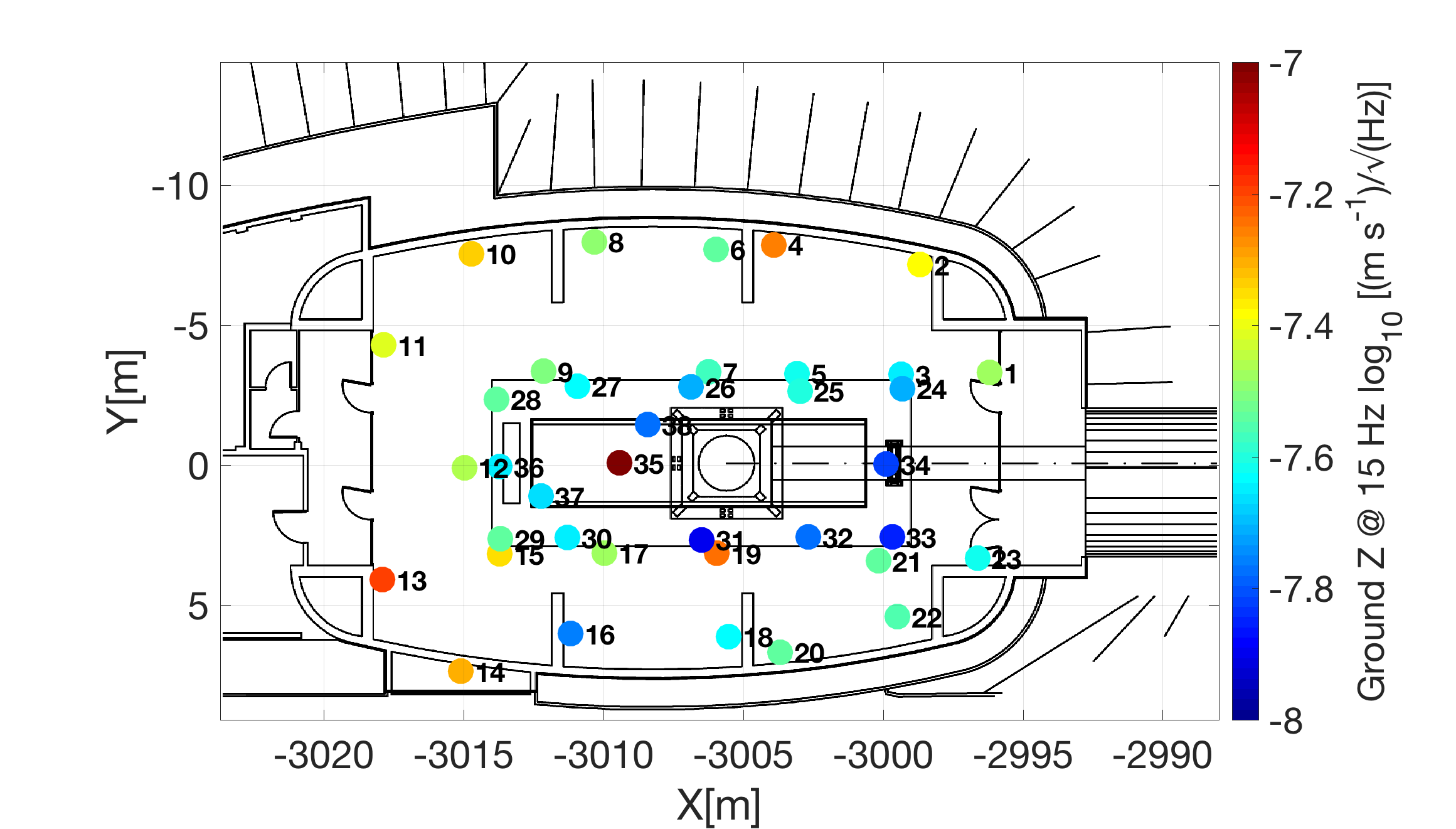}\\

\includegraphics [width=0.49\textwidth]{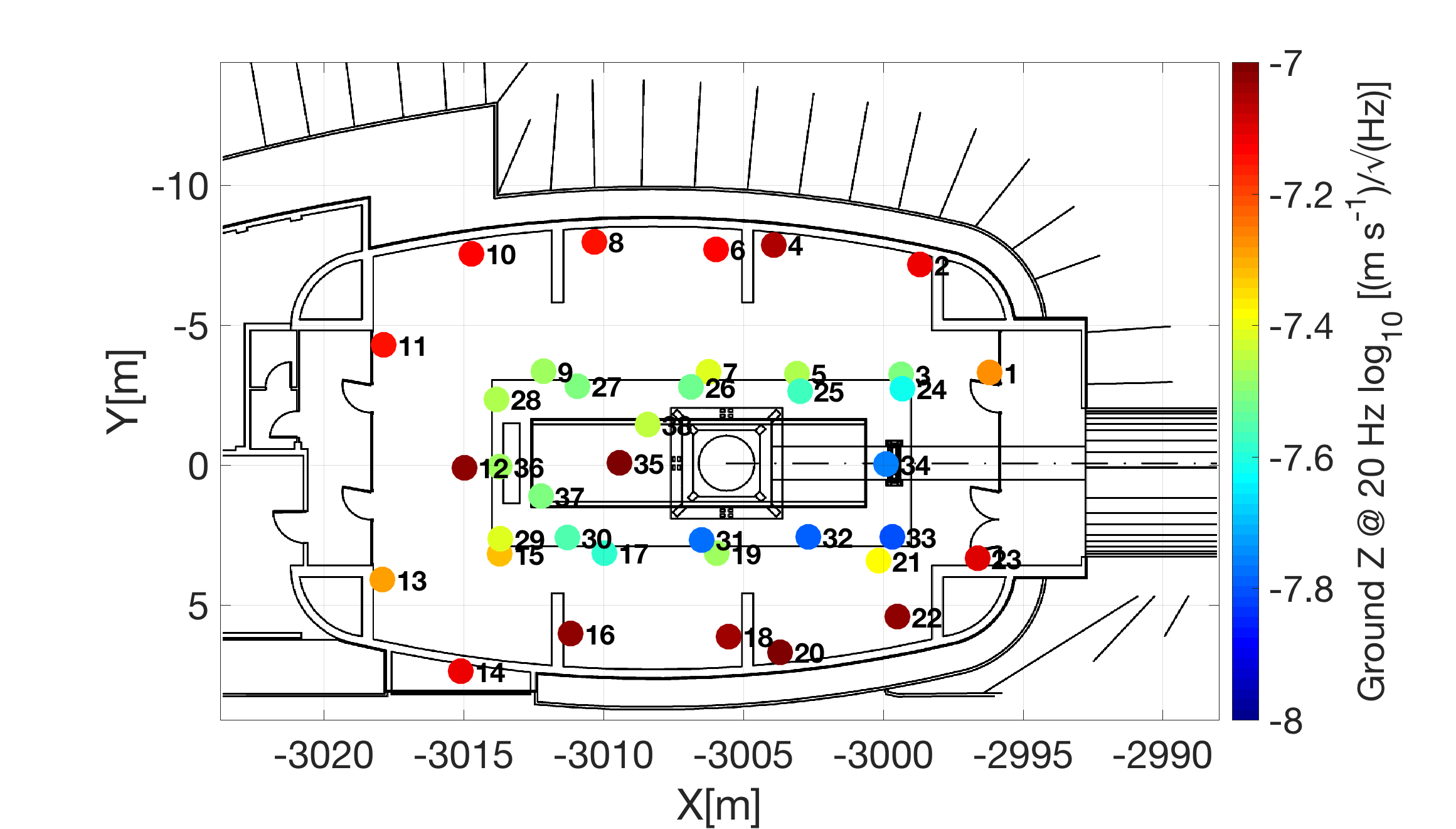}
\includegraphics [width=0.49\textwidth]{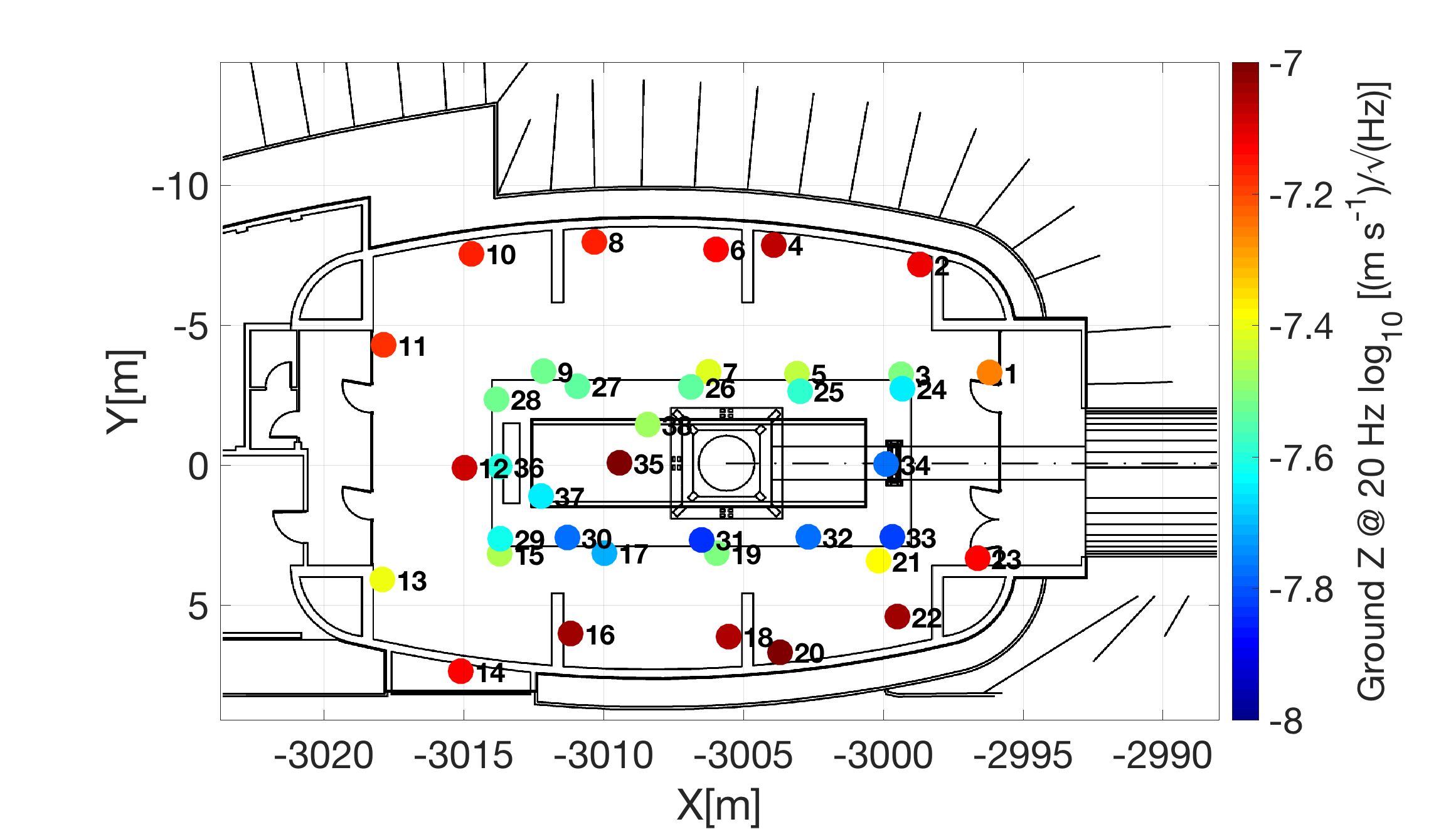}
\caption[]{Ground-vibration spectra at sensor locations. Both columns refer to hour 00:00:00 (UTC) of February 4 (left column) and February 5 (right column).  The three rows correspond to the frequencies 10, 15, 20\,Hz. We present the color coded values of the logarithm of the ASD at each location.}
\label{fig:dot_spectra}
\end{center}
\end{figure}

The array maps in Figure \ref{fig:dot_spectra} give an overview of the contribution to noise levels at frequency values 10\,Hz, 15\,Hz and 20\,Hz as measured in two different days. For some sensors, there can be significant differences between the spectra measured at the two days, but variations are generally small. Especially at 10\,Hz, one can see for both days that seismic motion is stronger towards negative $X$ values, which is consistent with seismic sources being located in the technical room. Focusing on sensors with $X>-3015\,$m, there is only a minor difference in spectra between sensors located on the building and tower platform (except for two sensors located close to the north wall). Looking at higher frequencies, a different behaviour was observed. At 15\,Hz, in addition to the disturbance by the technical room, major disturbances seem to originate from near the building walls and the vibrations propagate towards the tower platform. The origin of this noise is unclear, and even correlation measurements as shown below do not help to identify the sources. At 20\,Hz, there is a significant difference between spectra among the two slabs of the building. The sources of dominant ground vibrations seem to come from building walls. 

\begin{figure}[ht!]
\begin{center}
\subfigure[\scriptsize{Data 2018:02:04 h 00:00:00 (UTC)}]
{\includegraphics [width=4.0cm]{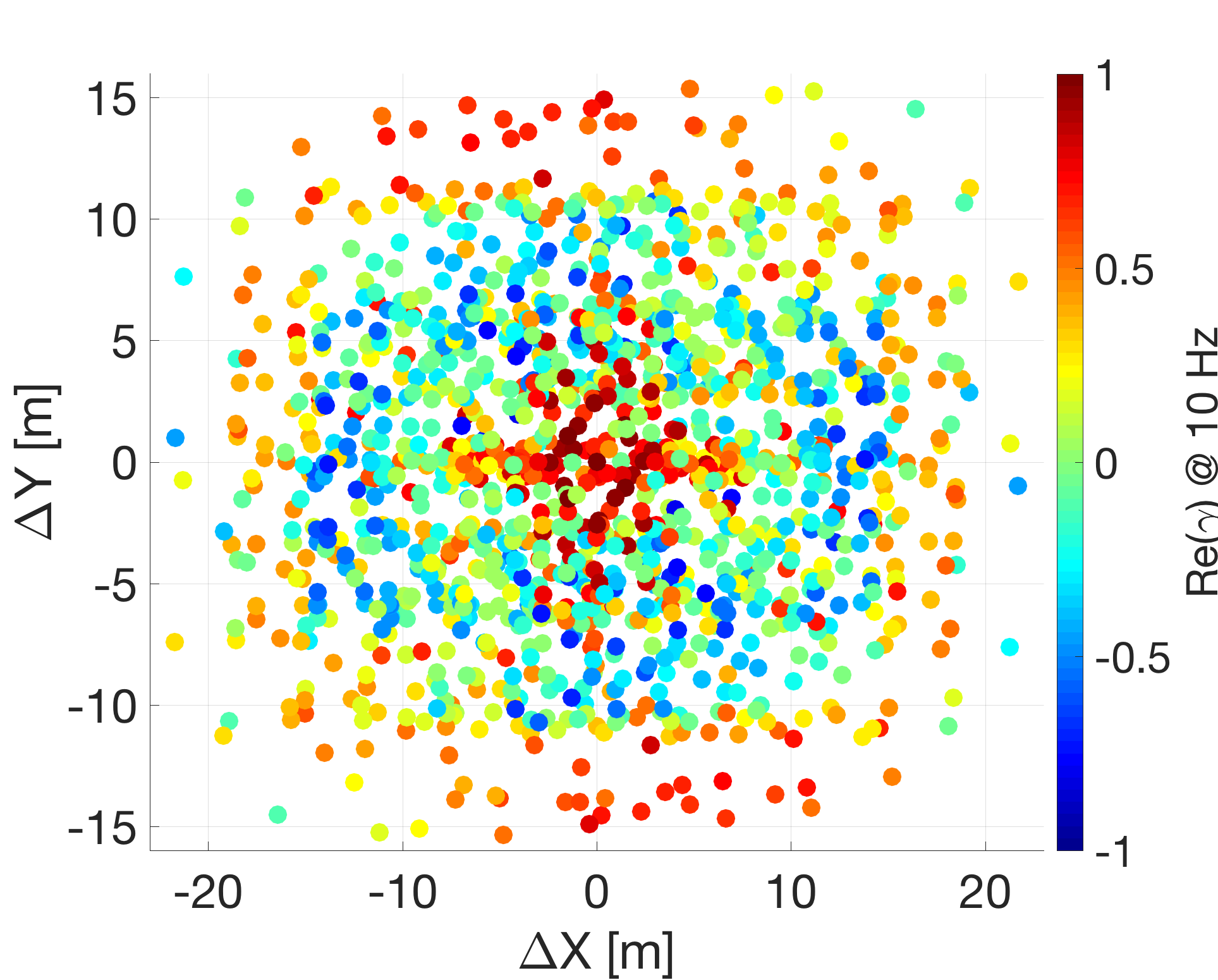}
\includegraphics [width=4.0cm]{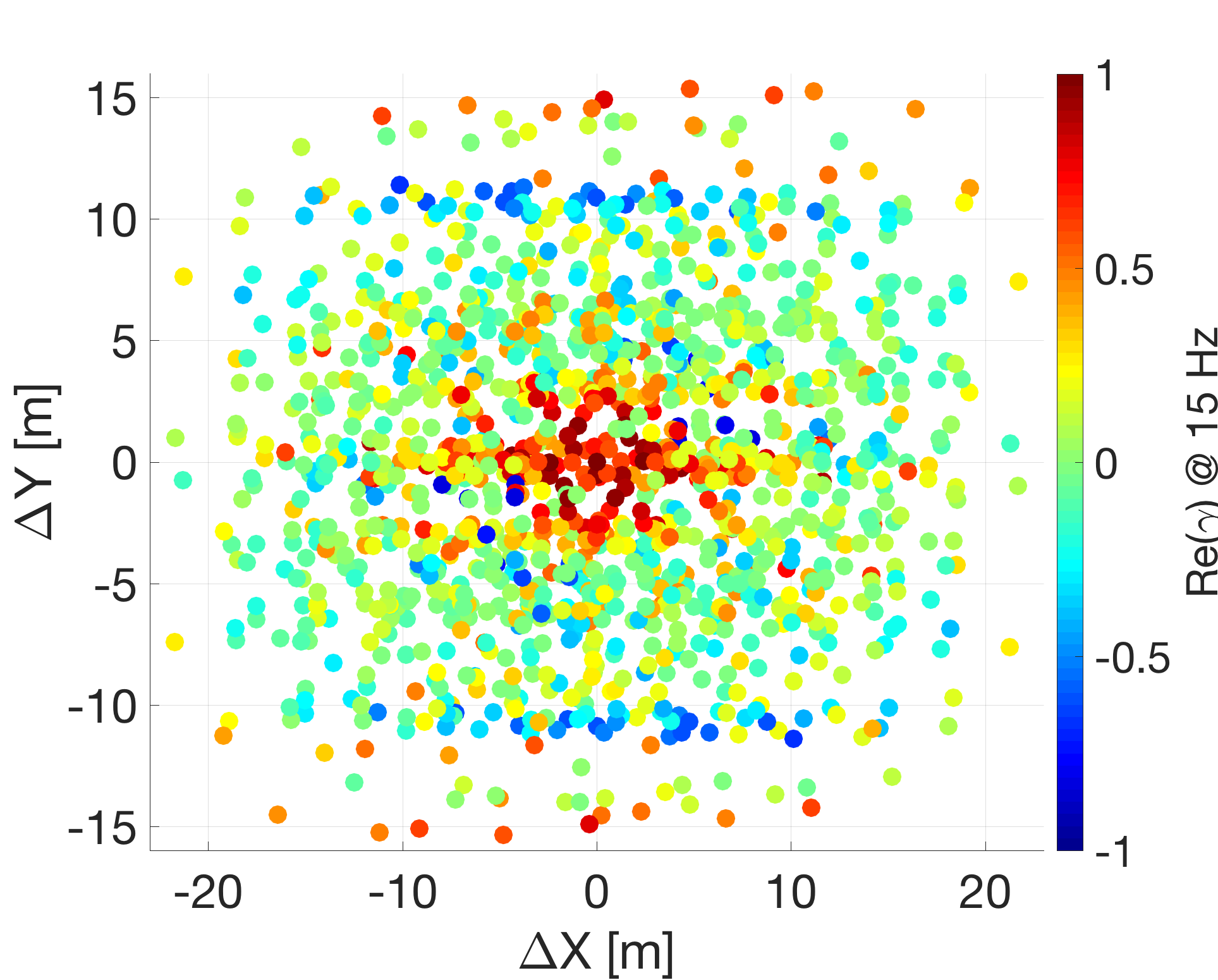}
\includegraphics [width=4.0cm]{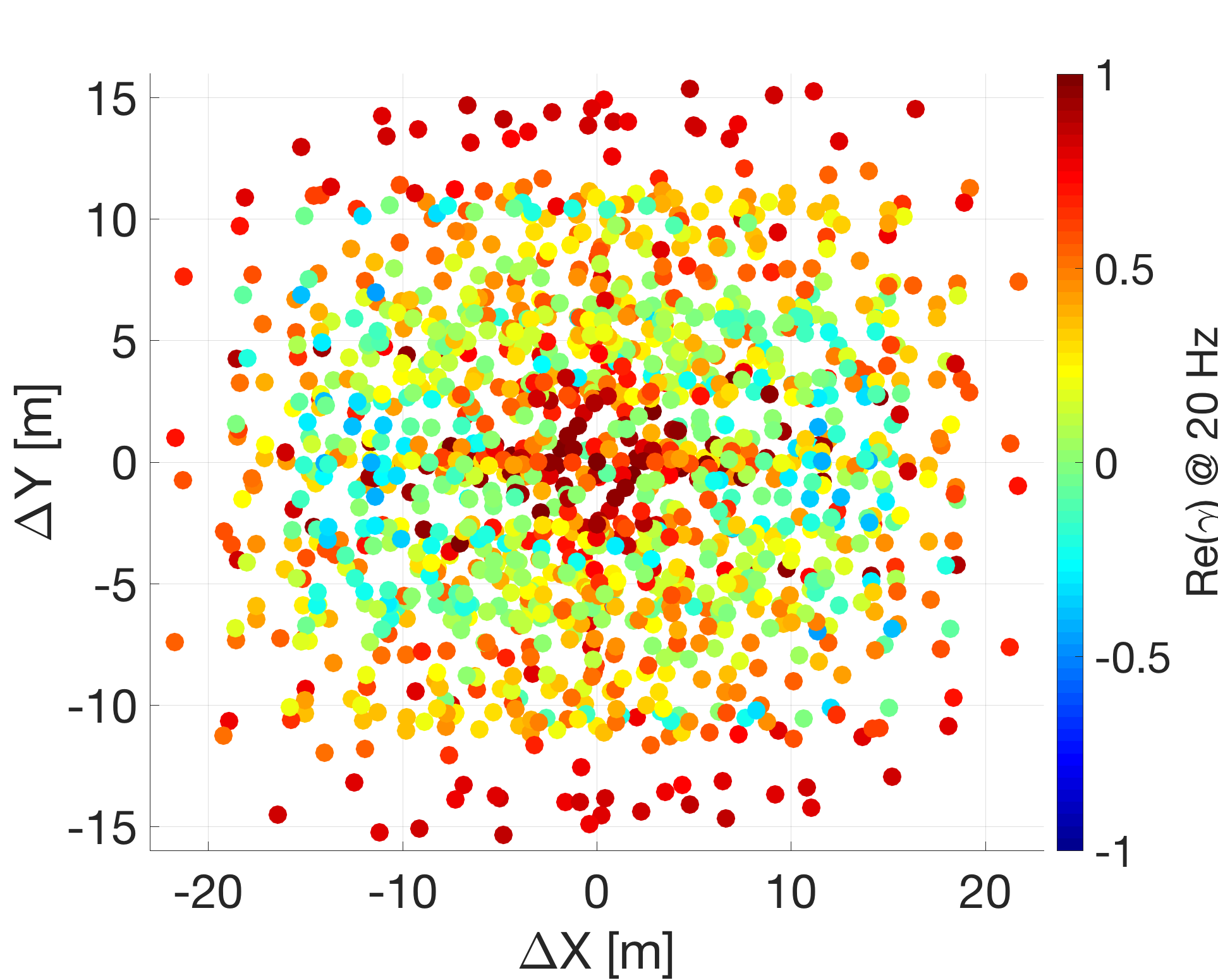}
}
\subfigure[\scriptsize{Data 2018:02:05 h 00:00:00 (UTC)}]
{\includegraphics [width=4.0cm]{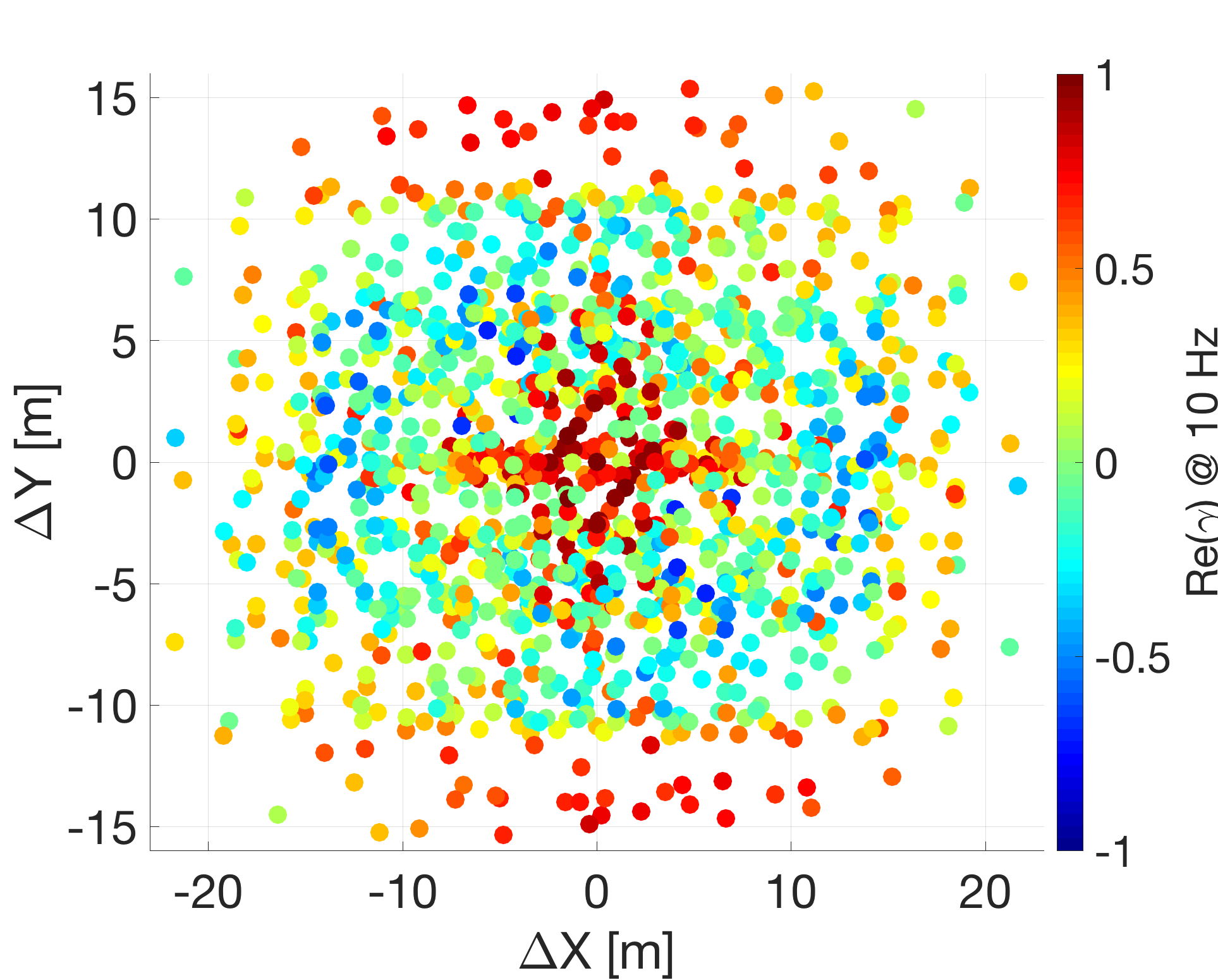}
\includegraphics [width=4.0cm]{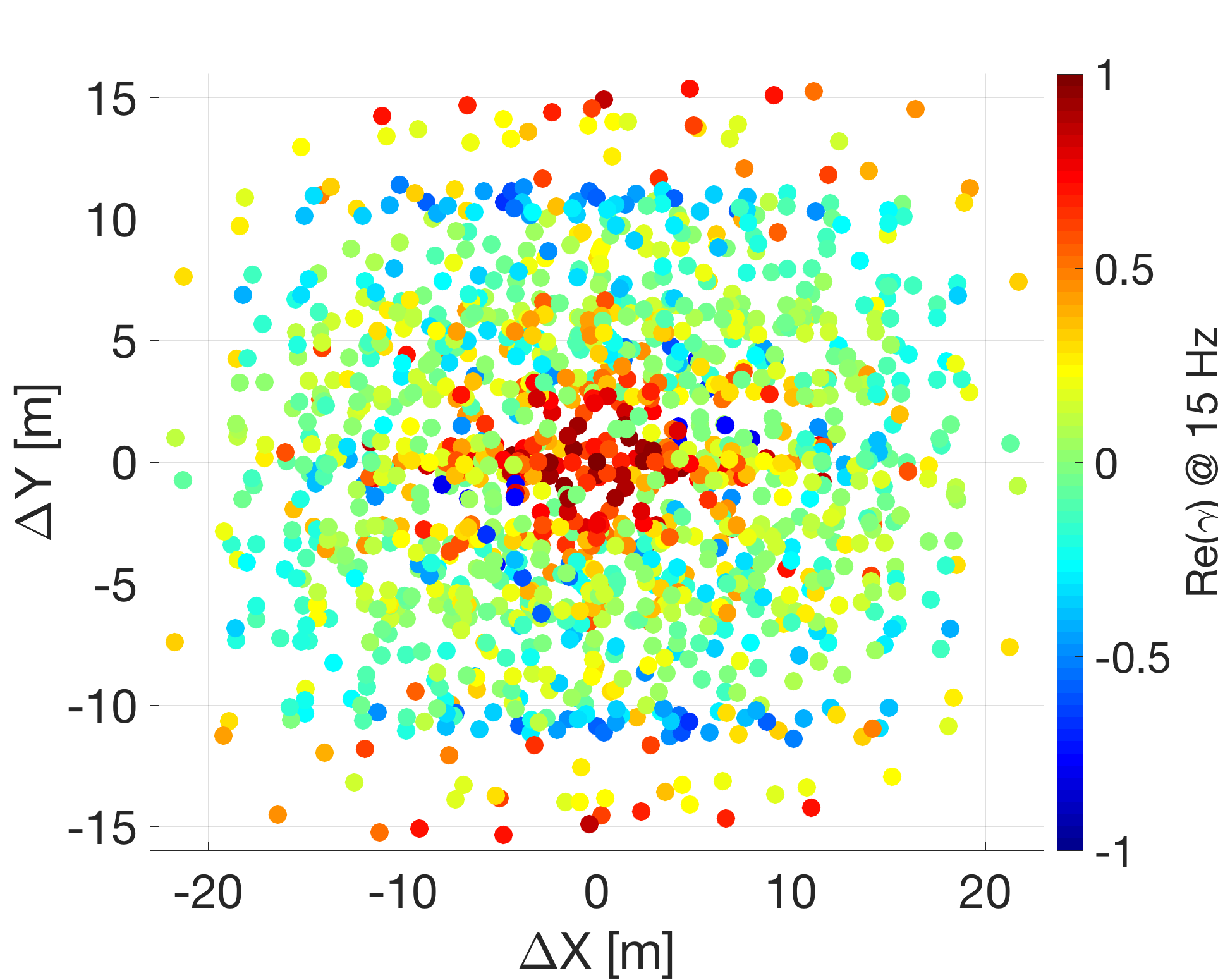}
\includegraphics [width=4.0cm]{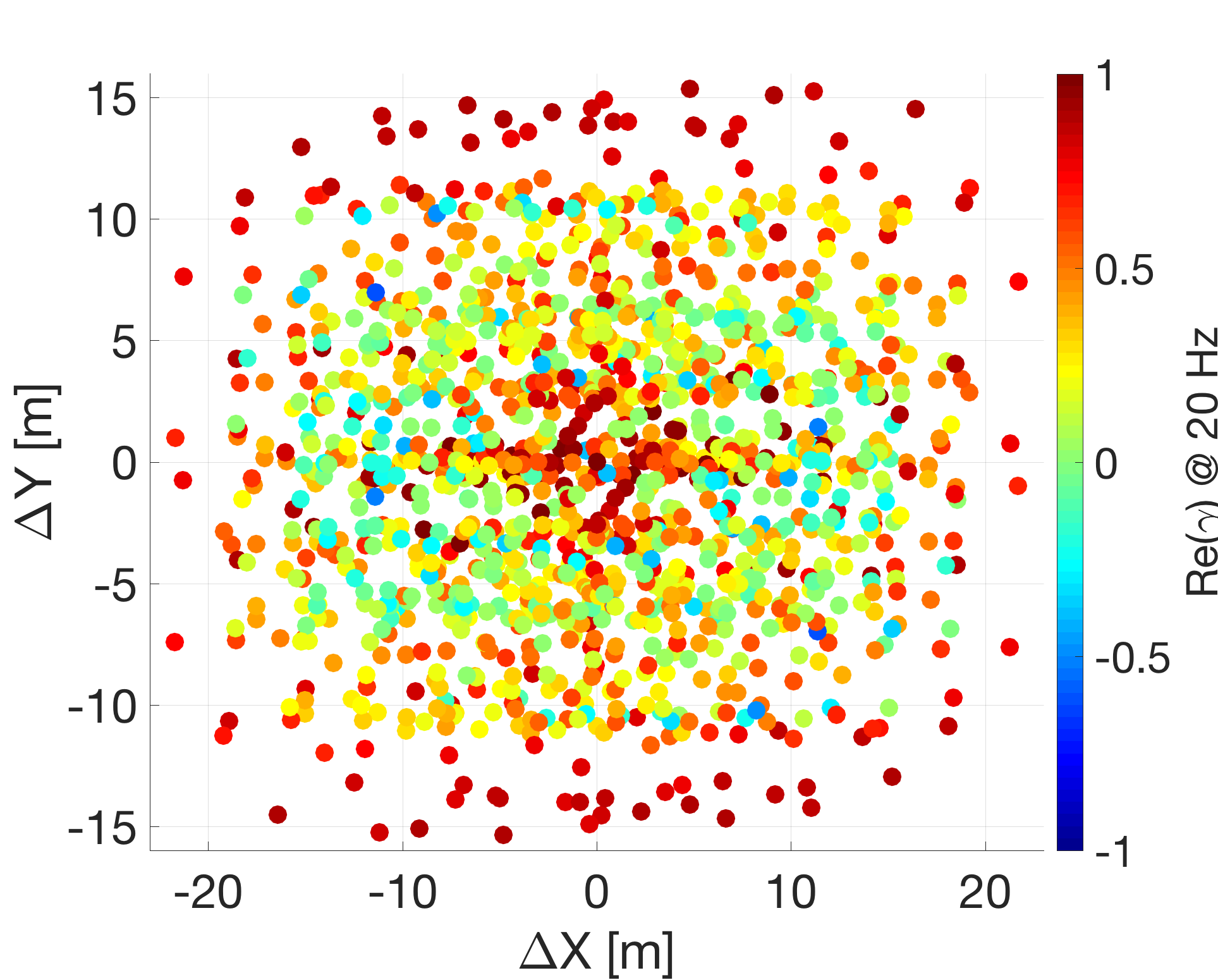}}
\caption[]{\small{Real part of $\gamma(f)$ between all seismometers at 10, 15 and 20 Hz. The coordinates x,y are relative position vector between two seismometers. Each pair of seismometers is plotted twice.}}
\label{fig:residual}
\end{center}
\end{figure}

The coherence $\gamma(f)$ between all seismometers in the array has also been analyzed. In general, the coherence between two signals $x(t)$ and $y(t)$ is defined as:
\begin{equation}
 \gamma(f) = \frac{S_{xy}(f)}{\sqrt{S_{x}(f)\cdot S_{y}(f)}} ,   
\end{equation}
where $S_{xy}(f)$ is the cross-spectral density between $x$ and $y$, $S_{x}(f)$ and $S_{y}(f)$ the power-spectral
densities of $x$ and $y$. The coherence is a complex-valued function. It plays a crucial role for a deeper understanding of the seismic field, for an accurate estimation of NN, and also for the design of a NN cancellation system \cite{Jan,Cel2000,Beker2011,CoEA2016a}. Sensor correlations define the Wiener filter (see Eq.(\ref{eq:WF_miso})). In Figure \ref{fig:residual} we present  the real part of coherence $\gamma(f)$ as a function of the relative position of seismometer pairs at three frequencies: 10, 15 and 20\,Hz. If the ground were homogeneous and isotropic, and Rayleigh waves dominated, we would expect the coherence evolving smoothly following a Bessel function \cite{Jenne}:
\beq
\gamma(f)=J_0(2\pi f r/c),
\eeq
where $r$ is the distance between sensors, and $c$ the speed of Rayleigh waves. It is real-valued, because the underlying model of the seismic field is isotropic. \cite{Jan2015,Jenne,Aki1957,Beker2010}. However, seismic correlations at the WEB are different: in Figure \ref{fig:residual}, points of noticeably different color can be next to each other (heterogeneity) and the correlation maps do not have polar symmetry (anisotropy).

Most likely the cause of the heterogeneity could be local seismic sources which affect some seismometers altering the spatial correlation throughout the array.
Still, at 10 Hz, ignoring cases of inhomogeneity, one can at least discern the underlying Bessel function, since seismic scattering from the tower platform is weak, and seismic sources are sufficiently far from all sensors.
At 15\,Hz, the correlation map assumes an $X,Y$ anisotropy supported by most sensor pairs, while the 20\,Hz map appears as an almost random pattern with high values of $\Re(\gamma)$ at almost any distance between seismometers. It is especially intriguing that ground vibrations at the north and south walls have high values of $\Re(\gamma)$, since Figure \ref{fig:dot_spectra} suggests that there are \emph{local} seismic sources near the walls leading to increased ground motion. We do not have a good explanation to offer. Potential explanations are: (1) sound waves inside the building push coherently on the north and south walls, (2) the increased ground vibration at the walls is not due to local sources, but due to some amplification of ground vibrations caused, for example, by a coupling between ground and walls. Both explanations certainly have their weaknesses, and a dedicated analysis needs to clarify this issue.

\subsection{Wiener filtering}
\label{sec:WF}

In the context of NN cancellation, the goal of Wiener filtering is to make an estimate of NN using data from a seismic array, and to subtract this estimate from the GW data. Since, nowadays, the NN lies still below the detector noise floor, we have tested the potential quality of the Wiener filtering by trying to reconstruct the signal from a target seismometer using the remaining seismometers as inputs to the Wiener filter. In this section, we first estimate the quality of the Wiener filtering as a function of the number of seismometers used as filter input. Then, we also investigate the optimal choice of seismometers, that is the optimum array layout, and eventually, for a given layout, we verify the stability of the Wiener filter with time by analyzing the stability of subtraction performance over a few hours of data.\\
The residual signal remaining after the subtraction of the Wiener filter output from the known target signal provides an estimate of the efficiency of Wiener filtering method. The average relative noise residual related to the original NN spectrum can be written as:

\begin{equation}
    R(f)= 1 -\frac{\vec{C}_{SX}^{ T}(f)\cdot \mathbf C_{SS}^{-1}(f)\cdot \vec{C}_{SX}(f)}{C_{XX}(f)},
    \label{eq:RES}
\end{equation}
where ${C}_{SX}$ denotes the vector of cross-spectral densities between the target sensor and seismometers, $\mathbf C_{SS}$ represents the matrix of cross-spectral densities between all seismometers in the array, and ${C}_{XX}$ is the power spectral density of the target sensor. In the plots and analysis below, we will show the square root of the relative residual $R$ to refer to the noise affecting signal amplitudes instead of power.

\begin{figure}[ht!]
\begin{center}
\includegraphics [width=0.7\textwidth]{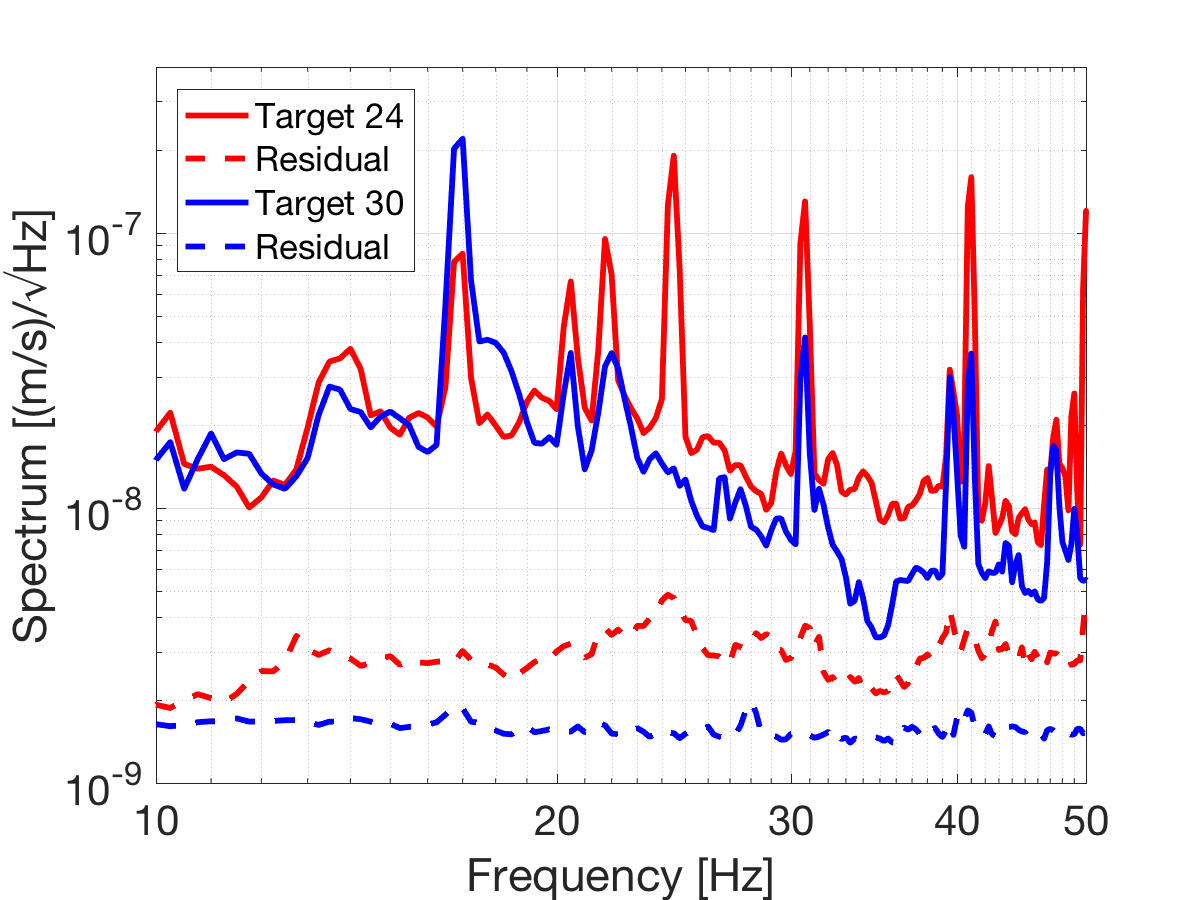}
\caption[]{Vertical ground motion at sensors 24 and 30, and corresponding absolute residuals after Wiener filtering with all other seismometers of the array used as input channels.}
\label{fig:residual_spectra}
\end{center}
\end{figure}

\begin{figure}[ht!]
\begin{center}
\includegraphics [width=0.49\textwidth]{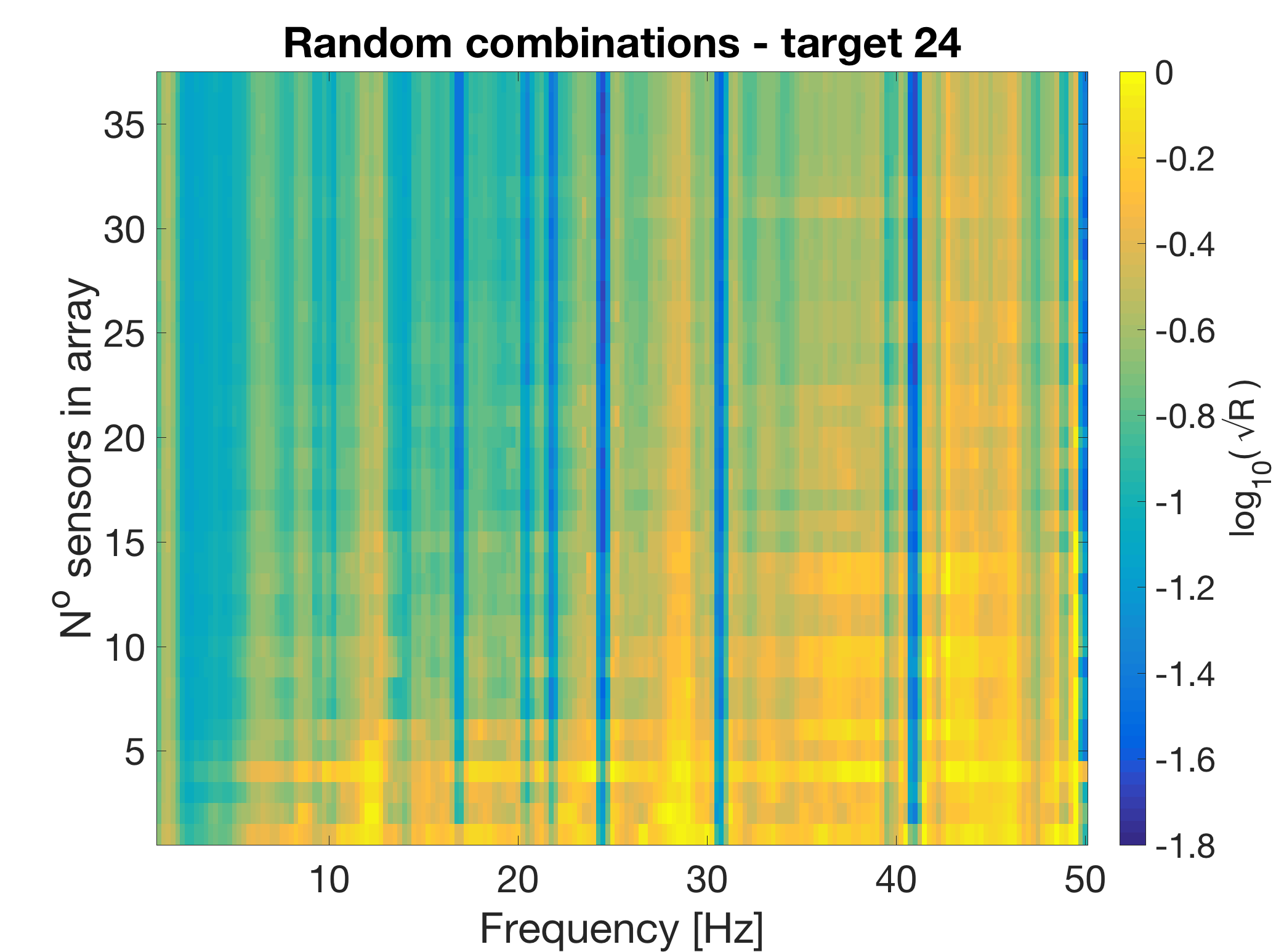}
\includegraphics [width=0.49\textwidth]{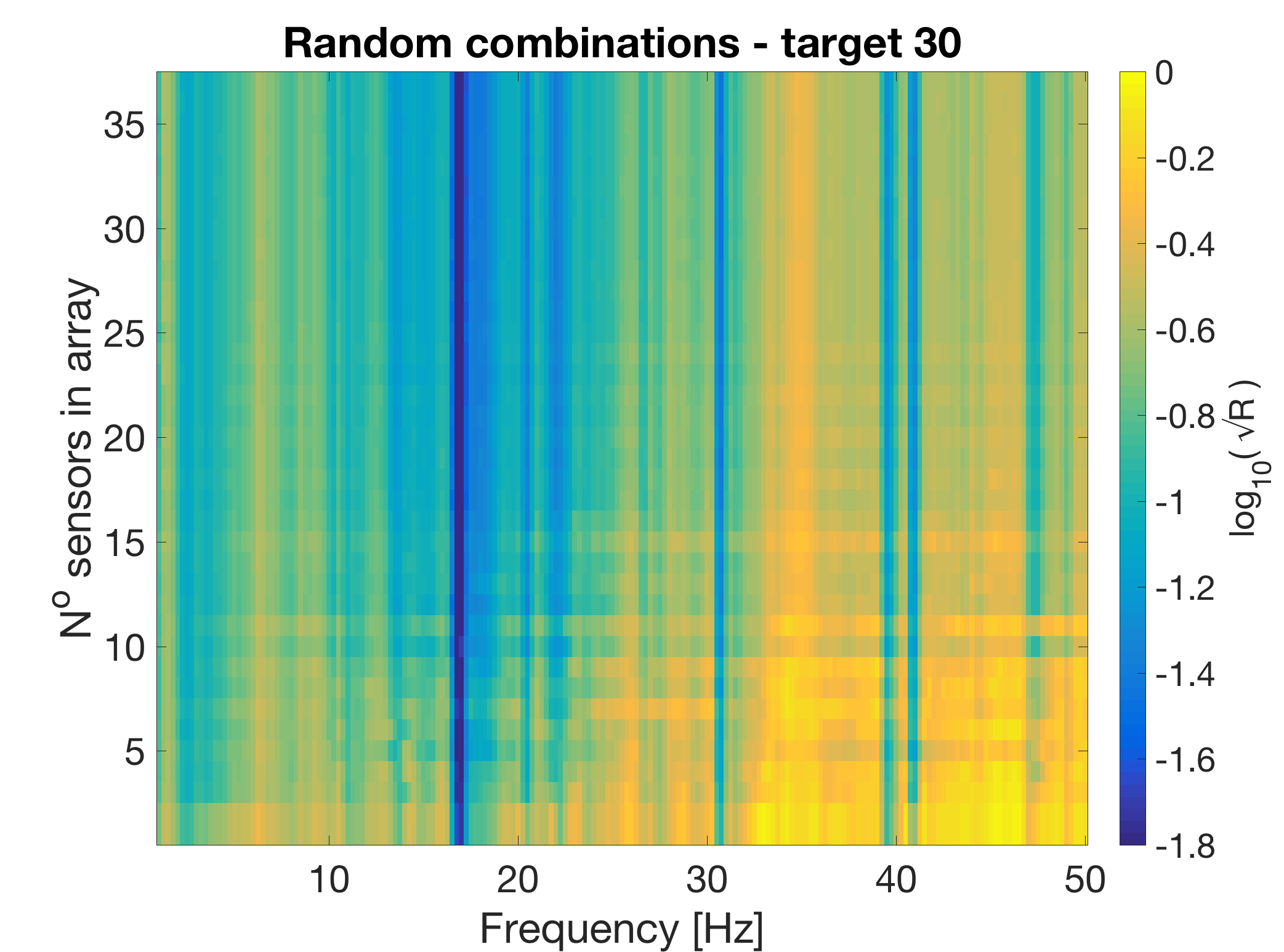}
\caption[]{Relative residuals for well-performing array configurations minimizing Eq.(\ref{eq:RES}) at 15\,Hz among 10 random selections of seismometers for each value of $N$. The residuals are computed for February 5, 00:00:00 (UTC) and for two different target sensors.}
\label{fig:all_residuals}
\end{center}
\end{figure}

In order to assess the quality of Wiener filtering as a function of the number and layout of the sensors, we chose to investigate two cases taking as targets sensor \#24 and \#30, both located on the tower platform. First, one can compare the spectra of the two target sensors and the absolute residual noise after cancellation with all other seismometers of the array used as input channel, as shown in Figure \ref{fig:residual_spectra}. It can be seen that the absolute residual spectra are relatively flat compared to the original spectra. The absolute residual of target sensor \#30 is about a factor~10 above the specified instrument self--noise \cite{arrayNikhef}. Flatness suggests that the residual is dominated by instrument self--noise which would mean that at least some of the seismometers in the array do not perform as well as expected. Further investigations are required.

For each number of seismometers picked from the total array as input channels, we selected ten random combinations covering the full possible range of $N=1,\ldots, 37$ of input channels. Among these random combinations, we chose the array with the best subtraction at 15\,Hz. We present the results of the analysis in Figure \ref{fig:all_residuals}. The plot shows the relative residuals $\sqrt{R}$ as function of the frequency for the best--performing arrays with the given number of sensors. The relative residuals are computed using the data from February 5 at midnight. The first thing to notice, consistent with Figure \ref{fig:residual_spectra}, is that the higher the original seismic spectra the better is the noise suppression factor, since the residual spectra are approximately flat. This means that the array captures all the necessary information about the seismic field, below 25\,Hz if more than about 6 sensors are used, above 25\,Hz if more than about 25 sensors are used. Also, the difference between the two plots in Figure \ref{fig:all_residuals} is mostly due to a difference between seismic spectra of sensors 24 and 30, and to a lesser extent a result of limited subtraction performance (again, provided that the number of input channels to the Wiener filter is sufficiently high). For both plots, an outlier is observed in relative residual values ($\sqrt{R}>1$). This behaviour can be related to a transient signal in the seismic data and for this reason the upper limit of the colour bar is set to zero. 
It should be noted that in some region of the graphs in Figure \ref{fig:all_residuals} the filtering seems to worsen when increasing the size of the reference array. This is due to the fact that we have not investigated all possible sensor sub--arrays but we limited ourselves to only ten random combinations for each case. Furthermore, picking the best performing array at 15\,Hz does not mean that it leads to the best performance at all frequencies. Thus the results presented in Figure \ref{fig:all_residuals} should be considered only as an indication of the Wiener--filtering quality.

\begin{table}[h]\footnotesize
\centering
\begin{tabular}{c l l }   
\hline
N$^{\circ}$ of seismometers& Target $\#$24 & Target $\#$30  \\
 in sub-array &  &   \\
\hline
1 & 25 & 29 \\
2 & 25, 26 & 29,31 \\
3 & 25, 26, 29 & 27, 29, 31  \\
4 & 25, 26, 30, 31& 26, 27, 29, 31 \\
5 & 25, 26, 30, 31, 35 & 24, 25, 27, 29, 31 \\
6 & 25, 26, 29, 30, 31, 35 & 24, 25, 27, 29, 31, 33 \\
7 & 25, 26, 27, 29, 30, 31, 35 & 24, 25, 26, 29, 31, 33, 35 \\
8 & 25, 26, 27, 29, 30, 31, 33, 35 & 24, 25, 26, 27, 29, 31, 33, 35\\
9 & 25, 26, 27, 29, 30, 31, 32, 33, 35 & 24, 25, 26, 27, 29, 31, 32, 33, 35 \\
\hline
\end{tabular}
\caption{\small{Optimal configurations at 15~Hz for target sensors 24 and 30. Sub--arrays are obtained by looping over all possibile configurations.}}
\label{tab:opt_config}
\end{table}

\begin{figure}[ht!]
\begin{center}
\includegraphics [width=\textwidth]{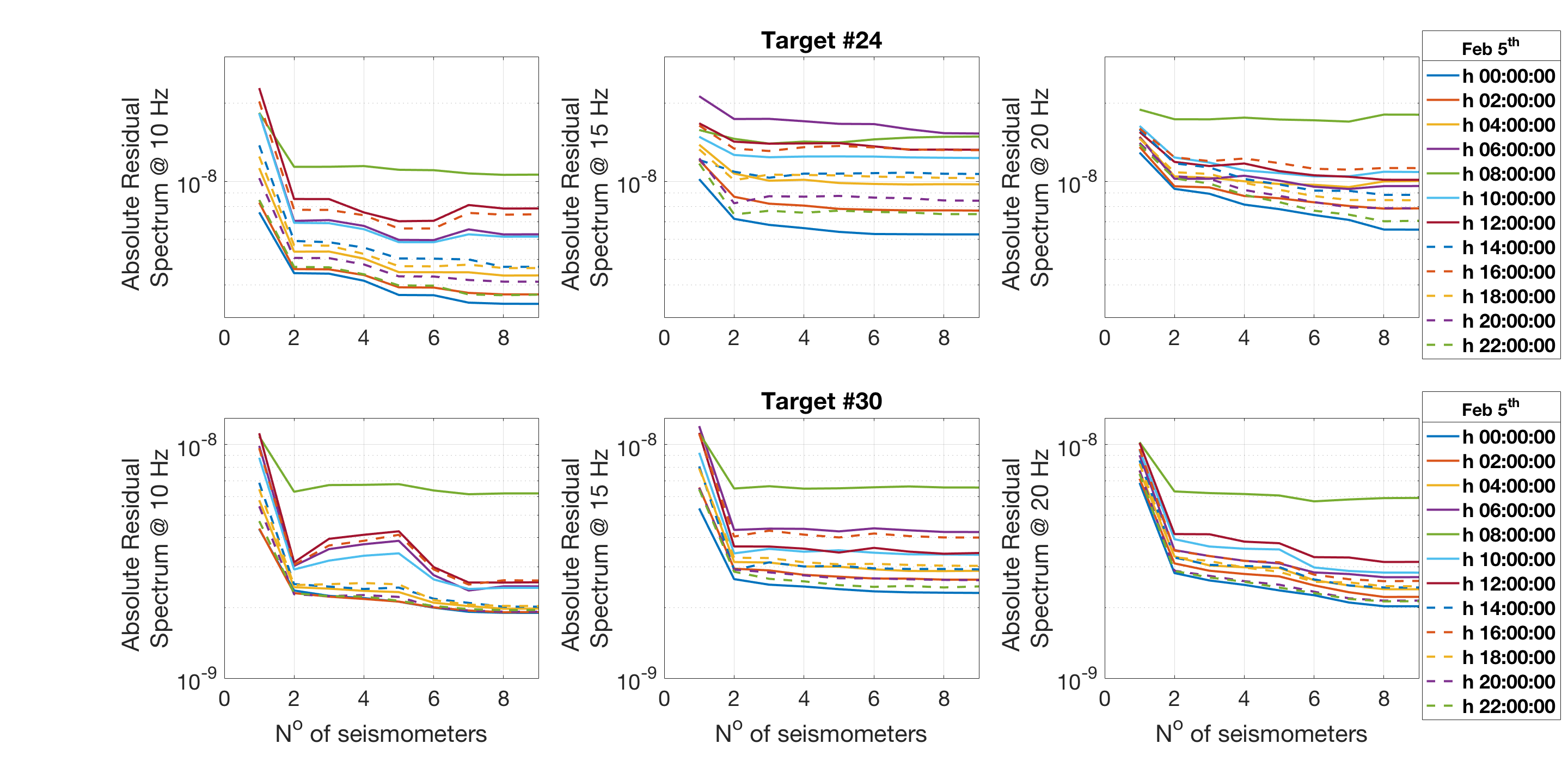}
\caption[]{Daily evolution of absolute residuals for well-performing input arrays on tower platform using sensors 24 (top row) and 30 (bottom row) as target channels. The $x$ and $y$-axes are the number of sensors and the square root of absolute residual in units
${\rm m{\cdot}s^{-1}/\sqrt{Hz}}$, respectively. The Wiener filter is calculated using one hour starting at midnight.}
\label{fig:fixcomb_1day}
\end{center}
\end{figure}

The second test is verifying how stable the Wiener filtering is over time.  For this purpose, we have calculated the Wiener filter using one hour of data starting at midnight and then we have applied it to the following hours to check if the residuals remained similar. We have analyzed the filter evolution using data from February 5 focusing on the seismometers located on the tower platform. We chose sensors \{24 25 26 27 29 30 31 32 33 35\} as input sensors, excluding sensors 24 and 30, which served as the target sensors. Best performing sub--arrays were selected, as in the previous study, based on the relative residual at 15\,Hz and they are reported in Table \ref{tab:opt_config}. In Figure \ref{fig:fixcomb_1day}, we present the absolute residual as function of the number of seismometers in the array for several hours of measurementduring the day of February 5, always using the same Wiener filter calculated at midnight. 

\begin{figure}[ht!]
\begin{center}
\includegraphics [width=0.49\textwidth]{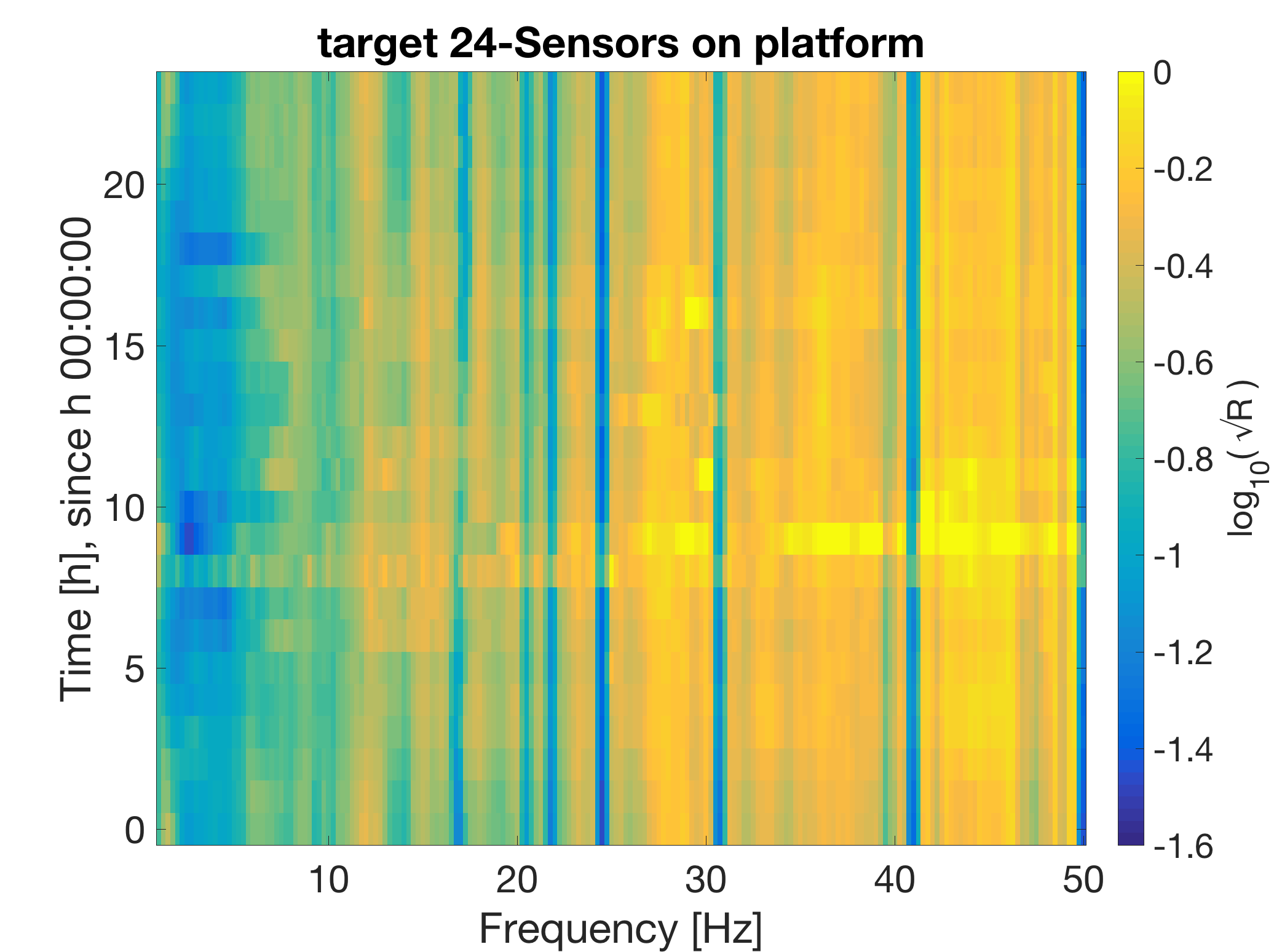}
\includegraphics [width=0.49\textwidth]{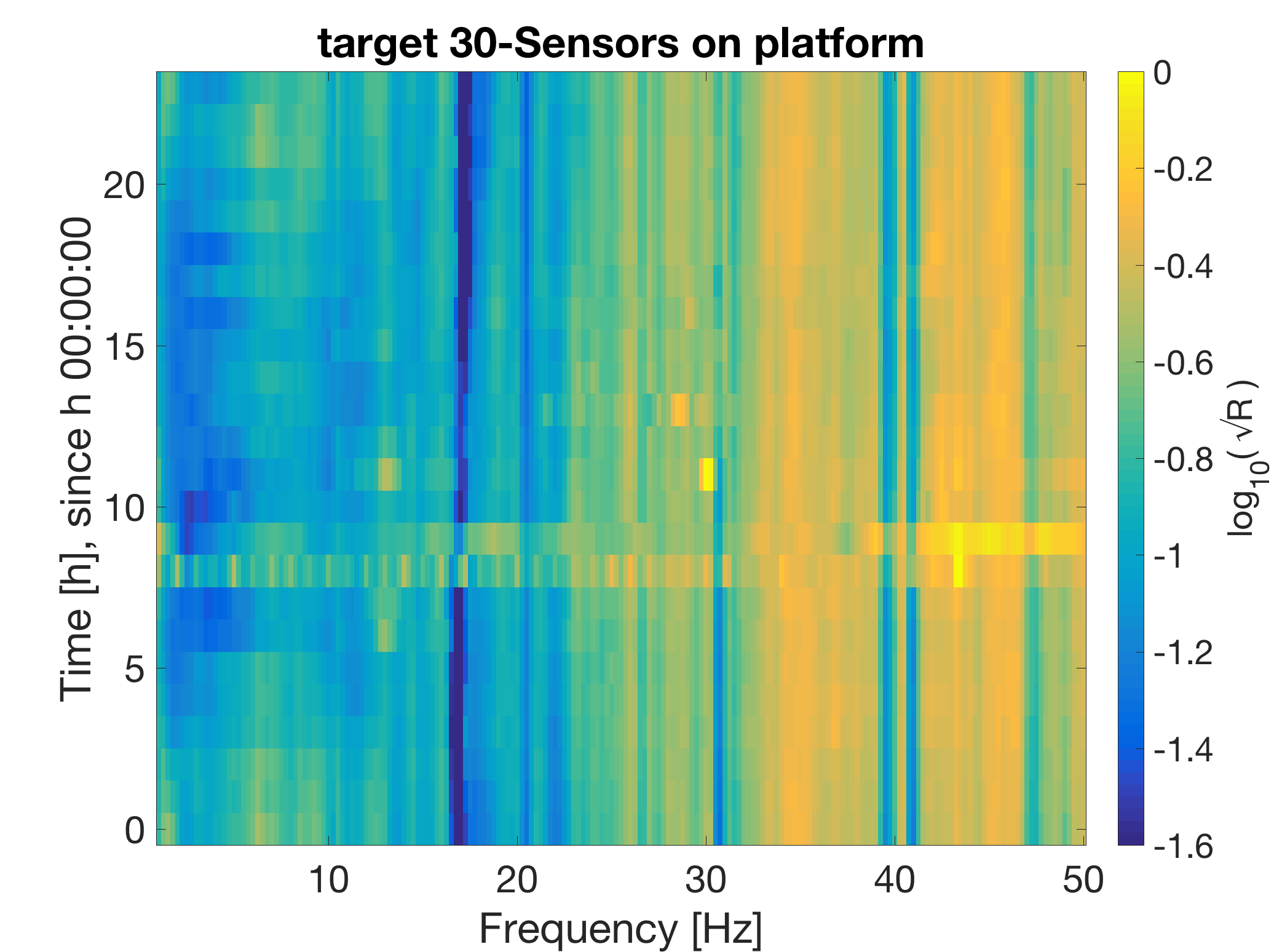}
\caption[]{\small{Time frequency map of square root of Wiener filter relative residuals for nine input sensors located on the tower platform. Data from February 5 is analyzed for target 24 (left plot) and 30 (right plot).}}
\label{fig:TF_wiener}
\end{center}
\end{figure}

At first glance, there is no trend in the performance degradation over time. For example, during a single hour, the absolute residual values are stable while increasing the number of seismometers in the sub-array. However, Wiener filter does not perform so well during the day noticing significant variations in performance.
We observe an increased absolute residual (solid green line), corresponding to the hour 08:00:00 (UTC), probably due to strong transients from human activity. However, the behavior of the absolute residuals is different from frequency to frequency. For example, at 15\,Hz and with sensor 24 as target, absolute residual values increase throughout the hours $\sim {\rm 6\cdot{10^{-9}m{\cdot}s^{-1}/\sqrt{Hz}}}$ at midnight to $\sim {\rm 1\cdot{10^{-8}m{\cdot}s^{-1}/\sqrt{Hz}}}$ at 16:00:00 (dashed red line) and then decrease to $\sim {\rm 8\cdot{10^{-9}m{\cdot}s^{-1}/\sqrt{Hz}}}$ at 22:00:00 (dashed green line). The significant variation over time of noise absolute residuals can be attributed to slow diurnal evolution of anthropogenic noise. This means that the Wiener filter will have to be updated at least every hour to ensure a stable background level.


The cancellation performance appears to be stable over time when looking  at the time evolution of the Wiener--filter relative residuals as a function of the frequency (see plots in Figure \ref{fig:TF_wiener} ). The noise residuals relative to the seismic input level do not vary significantly over time with the exception of one louder hour at 8~AM. 



\section{Conclusions} 
\label{sec:Conclusions}

In this paper, we presented a characterization of the seismic field at Virgo's West End Building required for the design of a Newtonian--noise cancellation system. The focus lay on the spectral and two--point spatial correlation features of the local ground motion, and on their impact on the performance of a Wiener filter for the cancellation of seismic signals. 

Correlation analysis has underlined the presence of inhomogeneities and anisotropies of the seismic field likely associated with local noise sources as well as with scattering of the seismic waves from the sub--surface structures of the building foundations. These effects become especially visible at higher frequencies, i.e., well above 10\,Hz, which is expected in presence of scattering. 

We have performed a case study of Wiener filtering using two different seismometers on the tower platform as target channels. We showed that the subtraction of the target signal can be efficiently achieved with at least six seismometers as input to the Wiener filter for frequencies below 25\,Hz, while more seismometers are required at higher frequencies. These figures are only relevant to the cancellation of seismic signals, not for Newtonian noise, but they allow us to estimate the level of correlations between different sensors. The results suggest that the array provides exhaustive information about the local seismic field for the purpose of Newtonian--noise cancellation at Virgo in the next future, since the observed correlations are sufficiently high for a noise suppression by up to a factor~5 (see Eq.(207) in \cite{Jan2015}). Moreover the high level of degeneracy observed in the array data suggests that a number of sensors lower than the 38~used in this work should be sufficient.\\
\indent The paper also shows evidence about how noise--cancellation performance varies over time when using a static Wiener filter. We have found a significant variation over time of absolute noise residuals that can be attributed to slow diurnal evolution of anthropogenic noise. The results suggest that the Wiener filter will have to be updated at least every hour to ensure a stable level of residual noise.

\ack{This research was supported by the TEAM/2016-3/19 grant from the Foundation for Polish Science and by the COST Action CA17137 (European Cooperation in Science and Technology).

\clearpage
\newpage
\bibliographystyle{unsrt}
\bibliography{bibliography}

\begin{thebibliography}{10}

\bibitem{Wei1972}
Rainer Weiss.
\newblock Electromagnetically coupled broadband gravitational antenna.
\newblock {\em MIT Res. Lab. Electron. Q. Prog. Rep}, 105:54--76, 1972.

\bibitem{Saulson}
Peter R.~Saulson.
\newblock Terrestrial gravitational noise on a gravitational wave antenna.
\newblock {\em Physical Review D - PHYS REV D}, 30:732--736, 08 1984.

\bibitem{Donatella}
Donatella Fiorucci, Jan Harms, Matteo Barsuglia, Irene Fiori, and Federico
  Paoletti.
\newblock Impact of infrasound atmospheric noise on gravity detectors used for
  astrophysical and geophysical applications.
\newblock {\em Phys. Rev. D}, 97:062003, Mar 2018.

\bibitem{Jan2015}
Jan Harms.
\newblock {Terrestrial Gravity Fluctuations}.
\newblock {\em Living Reviews in Relativity}, 18(3), 2015.

\bibitem{LSC2015}
{The LIGO Scientific Collaboration}, J~Aasi, B~P Abbott, R~Abbott, T~Abbott,
  M~R Abernathy, K~Ackley, C~Adams, T~Adams, P~Addesso, R~X Adhikari, V~Adya,
  C~Affeldt, N~Aggarwal, O~D Aguiar, A~Ain, P~Ajith, A~Alemic, et~al.
\newblock {Advanced LIGO}.
\newblock {\em Classical and Quantum Gravity}, 32(7):074001, 2015.

\bibitem{AcEA2015}
F~Acernese, M~Agathos, K~Agatsuma, D~Aisa, N~Allemandou, et~al.
\newblock {Advanced Virgo: a second-generation interferometric gravitational
  wave detector}.
\newblock {\em Classical and Quantum Gravity}, 32(2):024001, 2015.

\bibitem{Jan2016}
Jan Harms and Krishna Venkateswara.
\newblock {Newtonian-noise cancellation in large-scale interferometric GW
  detectors using seismic tiltmeters}.
\newblock {\em Classical and Quantum Gravity}, 33(23):234001, 2016.

\bibitem{Cel2000}
G.~Cella.
\newblock {Off-Line Subtraction of Seismic Newtonian Noise}.
\newblock In B.~Casciaro, D.~Fortunato, M.~Francaviglia, and A.~Masiello,
  editors, {\em Recent Developments in General Relativity}, pages 495--503.
  Springer Milan, 2000.

\bibitem{coughlin2018}
M.~W. Coughlin, J.~Harms, J.~Driggers, D.~J. McManus, N.~Mukund, M.~P. Ross,
  B.~J.~J. Slagmolen, and K.~Venkateswara.
\newblock {Implications of Dedicated Seismometer Measurements on
  Newtonian-Noise Cancellation for Advanced LIGO}.
\newblock {\em Phys. Rev. Lett.}, 121:221104, Nov 2018.

\bibitem{coughlin2014}
M~Coughlin, J~Harms, N~Christensen, V~Dergachev, R~DeSalvo, S~Kandhasamy, and
  V~Mandic.
\newblock {Wiener filtering with a seismic underground array at the Sanford
  Underground Research Facility}.
\newblock {\em Classical and Quantum Gravity}, 31(21):215003, 2014.

\bibitem{book1}
Jacob Benesty, M~Mohan Sondhi, and Yiteng Huang.
\newblock {\em Springer handbook of speech processing}.
\newblock Springer, 2007.

\bibitem{book2}
Saeed~V Vaseghi.
\newblock {\em Advanced digital signal processing and noise reduction}.
\newblock John Wiley \& Sons, 2008.

\bibitem{Virgo}
T.~Accadia et~al.
\newblock {Virgo: a laser interferometer to detect gravitational waves}.
\newblock {\em Journal of Instrumentation}, 7(03):P03012--P03012, mar 2012.

\bibitem{Paoli}
M.~Marsella et~al.
\newblock {Virgo Reference System (VRS) Control Point Network Monographs}.
\newblock {\em Technical report, VIR-0523Y-13}, 2018.

\bibitem{arrayNikhef}
M.~Beker et~al.
\newblock {Innovations in seismic sensors driven by the search for
  gravitational waves}.
\newblock {\em The Leading Edge}, 35(7):590--593, 2016.

\bibitem{data_indoor}
Indoor data.
\newblock $http://foka.ise.pw.edu.pl/virgo/NN_Data_Reparsed/$.

\bibitem{Jan}
Jan Harms, Bram J.~J. Slagmolen, Rana~X. Adhikari, M.~Coleman Miller, Matthew
  Evans, Yanbei Chen, Holger M\"uller, and Masaki Ando.
\newblock Low-frequency terrestrial gravitational-wave detectors.
\newblock {\em Phys. Rev. D}, 88:122003, Dec 2013.

\bibitem{Beker2011}
M.G. Beker, G.~Cella, R.~DeSalvo, M.~Doets, H.~Grote, J.~Harms, E.~Hennes,
  V.~Mandic, D.S. Rabeling, J.F.J. van~den Brand, and C.M. van Leeuwen.
\newblock {Improving the sensitivity of future GW observatories in the 1 - 10Hz
  band: Newtonian and seismic noise}.
\newblock {\em General Relativity and Gravitation}, 43(2):623--656, 2011.

\bibitem{CoEA2016a}
M~Coughlin, N~Mukund, J~Harms, J~Driggers, R~Adhikari, and S~Mitra.
\newblock {Towards a first design of a Newtonian-noise cancellation system for
  Advanced LIGO}.
\newblock {\em Classical and Quantum Gravity}, 33(24):244001, 2016.

\bibitem{Jenne}
Jennifer~C. Driggers, Jan Harms, and Rana~X. Adhikari.
\newblock {Subtraction of Newtonian noise using optimized sensor arrays}.
\newblock {\em Phys. Rev. D}, 86:102001, Nov 2012.

\bibitem{Aki1957}
Keiiti Aki.
\newblock Space and time spectra of stationary stochastic waves, with special
  reference to microtremors.
\newblock {\em Bull. Earthq. Res. Inst.}, 35:415--456, 1957.

\bibitem{Beker2010}
M.G. Beker, J.F.J. van~den Brand, E.~Hennes, and D.S. Rabeling.
\newblock {Towards time domain finite element analysis of gravity gradient
  noise}.
\newblock {\em Journal of Physics: Conference Series}, 228(1):012034, 2010.

\end{thebibliography}

\end{document}